\documentclass[]{elsarticle}
\usepackage[utf8]{inputenc}

\usepackage{grffile}
\usepackage{lineno}
\usepackage{hyperref}
\usepackage{times}
\usepackage{calc}
\usepackage{graphics}
\usepackage{epsfig}
\usepackage{amssymb}
\usepackage{amsmath}
\usepackage{flafter}
\usepackage{float}
\usepackage{latexsym}
\usepackage{psfrag}
\usepackage{dsfont}
\usepackage{rotating}
\usepackage{longtable}
\usepackage{cleveref}
\usepackage{color}
\usepackage{longtable,lscape}
\usepackage{multirow}
\usepackage{tabularx}
\usepackage{booktabs}
\usepackage{subcaption}
\usepackage{array}
\usepackage{soul}
\newcolumntype{M}[1]{>{\centering\arraybackslash}m{#1}}
\usepackage[nice]{nicefrac}
\usepackage{pgfplots}
\usepackage{tikzscale}
\usetikzlibrary{external}
\usepackage{pgfplots}
\usepgfplotslibrary{groupplots}
\pgfplotsset{
	compat=newest,
	legend image code/.code={
		\draw[mark repeat=2,mark phase=2]
		plot coordinates {
			(0cm,0cm)
			(0.2cm,0cm)        
			(0.4cm,0cm)         
		};%
	}
}
\usepackage{pgfkeys}
\usepackage{pgfplotstable}
\usetikzlibrary{matrix}
\usepgfplotslibrary{groupplots}
\pgfplotsset{compat=newest}
\pgfplotsset{}
\pgfplotsset{
	tick label style={font=\footnotesize},
	label style={font=\footnotesize},
	legend style={font=\footnotesize}
}
\pgfplotsset{major grid style={loosely dotted, thin, gray}}
\pgfkeys{
	/includetikzgraphic/.is family, /includetikzgraphic,
	default/.style = {
		width = 0.85\textwidth,
		ratio = 0.85},
	width/.estore in = \itgWidth,
	ratio/.estore in = \itgRatio
}
\newlength\figurewidth
\newlength\figureheight

\modulolinenumbers[5]

\journal{Journal of \LaTeX\ Templates}

\renewcommand{\vec}[1]{\mathbf{#1}}

\LetLtxMacro{\originaleqref}{\eqref} 
\renewcommand{\eqref}{equation~\originaleqref}

\begin{document}
	
\begin{frontmatter}

\title{A Perspective on Machine Learning Methods in Turbulence Modelling}

\author{Andrea Beck}
\author{Marius Kurz}

\address{Lab. of Fluid Dynamics and Technical Flows, Univ. of Magdeburg "Otto von Guericke", Germany}
\address{Institute of Aerodynamics and Gas Dynamics, University of Stuttgart, Germany}

\begin{abstract}
This work presents a review of the current state of research in data-driven turbulence closure modeling. It offers a perspective on the challenges and open issues, but also on the advantages and promises of machine learning methods applied to parameter estimation, model identification, closure term reconstruction and beyond, mostly from the perspective of Large Eddy Simulation and related techniques. We stress that consistency of the training data, the model, the underlying physics and the discretization is a key issue that needs to be considered for a successful ML-augmented modeling strategy. In order to make the discussion useful for non-experts in either field,  we introduce both the modeling problem in turbulence as well as the prominent ML paradigms and methods in a concise and self-consistent manner. Following, we present a survey of the current data-driven model concepts and methods, highlight important developments and put them into the context of the discussed challenges. 
\end{abstract}
\end{frontmatter}

\section{Introduction}\label{sec:intro}
With the recently exploding interest in machine learning (ML), artificial intelligence (AI), Big Data and generally data-driven methods for a wide range of applications, it is not surprising that these approaches have also found their way into the field of fluid mechanics~\cite{brunton2020machine}, for example for optimization and control~\cite{tang2020robust}, flow field reconstruction from experimental data~\cite{lee2017piv} or shock capturing for numerical methods~\cite{ray2018artificial,shock}. An area with a particularly high number of recently published research that uses ML methods is turbulence, its simulation and its modeling. The reasons for this marriage between these two fields are numerous, and some will be explored in this article. The most natural one might be - mostly from the perspective of computational fluid dynamics (CFD), but also from the experimentalist's view - is that when simulating or measuring turbulence, one naturally encounters large amounts of high-dimensional data from which to extract rather low-dimensional information, for example the drag coefficient of an immersed body. This generation of knowledge (or at least models and ideas about knowledge) from a large body of data is at the core of both understanding turbulence as well as machine learning - and of course many other fields. At least from the authors' perspective, there often seems to exist a mismatch between the tremendous amount of data from simulations or measurements that we have on turbulence, and the tangible results we can actually deduce from that. Here is where ML and Big Data might help - for at least in theory, they can  make more effective use of the data we have. \\
In this article, we intend to give an overview of recent research efforts into turbulence modeling with machine learning methods, or more precisely data-driven closure models and methods. It is not meant to replace the already existing valuable review articles on this matter, e.g.~\cite{duraisamy2019turbulence,pandey2020perspective,pollard2016whither}, but to provide a complementary perspective with a particular focus on Large Eddy Simulation (LES) methods, although many of the challenges and chances of ML in this field also directly transfer to the Reynolds Averaged Navier-Stokes (RANS) system. We have attempted to present both the simulative side and the ML side in a concise and self-contained manner with a focus on the basic principles, so that the text may be useful to non-specialists as well and provide a first starting point from which to venture further. The focus of this review is thus not so much on specific examples of how ML methods can augment turbulence modeling (a very daunting task given the current surge in publications), but on the challenges, chances and capabilities of these methods.\\ However, before discussing the combination of ML and turbulence models, we start in Sec.~\ref{subsec:turb} with a brief introduction to turbulence and the two prevalent simulation strategies, namely the RANS approach and the LES technique. The discussion is focused on the formal aspects of defining the governing equations and the associated solution, and on the factors that influence them. This is somewhat of a shift from the usual introduction to LES, but in the light of data-driven modeling, \emph{data-consistency} (during training) and \emph{model-data-consistency}  (during inference) are important aspects. Highlighting the different factors that influence the LES solution and its modeling is not only helpful here, but also motivates the use of ML methods in this complex, high-dimensional and non-linear situation. With this introduction in place, we will then give a perspective on the challenges and possibilities of ML and turbulence models in Sec.~\ref{subsec:why}. Here, we discuss what types of challenges arise in turbulence modeling which are mostly specific to the ML approach, but also give reasons to be optimistic that data-driven turbulence simulations can at least augment models and simulation approaches derived from first principles. Before discussing the recent literature and presenting concrete examples, we first review the most basic machine learning paradigms and methods in Sec.~\ref{sec:ml}. Here, as in the rest of the manuscript, we mostly focus on \emph{supervised} learning approaches. Also, since neural networks are the most commonly used method in this context, the discussion will also lean towards them and related methods. In Sec.\ref{sec:turb}, we present a possible hierarchy of modeling approaches, from data-based parameter estimation to the full replacement of the original partial differential equation (PDE) by an ML method. In this section, the focus lies on the general idea behind the methods, and on their relationship to the challenges and chances identified in Sec.~\ref{subsec:why}. Due to the dynamic research landscape and the high frequency of new publications and ideas, no converged state of accepted best practices has emerged yet, so our goal here is not to rank the proposed methods but to equip the reader with the conceptual knowledge of the current state of the art. Sec.~\ref{sec:outro} concludes with some comments and ideas for future developments.

\subsection{Turbulence and turbulence modeling}\label{subsec:turb}
Turbulence is a state of fluid motion that is prevalent in nature and in many technical applications. The blood stream in the bodies of mammals can become turbulent, and the resulting increased stresses can lead to the onset of serious medical conditions. Transition and turbulence govern the flow through turbo-machines and significantly influence their efficiency and reliability. Finally, turbulent drag and heat transfer not only strongly influence the fuel consumption of subsonic passenger aircraft, but also pose a great safety challenge for supersonic vehicles. This list is of course far from exhaustive, but serves to highlight the importance of understanding, analyzing and predicting turbulent flows. \\
As opposed to its opposite, the laminar flow state, turbulence is characterized by a broad range of interacting temporal and spatial scales. This is an expression and a result of the non-linearity of the governing equations which leads to rich dynamics and a high sensitivity to initial and boundary conditions. Turbulence shares this feature with other chaotic systems, and they all have to cope with a number of challenges in experimental investigations, mathematical analysis and also numerical approximations and models. Turbulence is typically characterized by the so-called Reynolds ($Re$) number, a dimensionless ratio of the actions of inertia and the counteracting viscous effects. The viscous 1D Burgers' equation for a variable $u(x,t)$ representing a velocity serves as an incomplete but useful model for the turbulent cascade and highlights the influence of the $Re$ number:
\begin{equation}\label{eq:burgers}
\frac{\partial u}{\partial t}+\frac{1}{2}\frac{\partial u^2}{\partial x}-\frac{1}{{Re}}\frac{\partial^2 u}{\partial x^2}=0.
\end{equation}
Assuming an initial solution at time $t=0$ with amplitude $\hat{u}$ and a single wavenumber $k$ as $u(x,0)=\hat{u}\,sin(kx)$, the initial time derivative becomes
\begin{equation}\label{eq:burgersdt}
\frac{\partial u}{\partial t}\bigg \vert_{t=0}=-\frac{1}{2}\hat{u}^2\,sin(2kx)-\frac{\hat{u}}{{Re}}k^2 sin(kx).
\end{equation}
Eq.~\ref{eq:burgersdt} already reveals a number of interesting insights. First of all, it is the non-linear convective term that drives the scale cascade by producing higher frequency contributions ($2kx$), while the viscous term damps the initial waveform ($kx$). The magnitude of this damping is a function of $k^2/Re$, i.e. regardless of $Re$, this term will dominate in the limit $k\rightarrow\infty$\footnote{The inviscid counterpart to the Navier-Stokes equations, the Euler equations, lack the limiting viscosity. There are some suggestions that the Euler system might develop singularities  and thus contain different dynamics than the Navier-Stokes equations, but this is not considered here.}, a fact that motivates the choice of dissipative closure models as will be discussed later. For small $k$ and/or large $Re$ however, the evolution of Eq.~\ref{eq:burgers} is dominated by the inviscid dynamics, while the dissipation only occurs at the smallest scales. For these cases,  Eq.~\ref{eq:burgers} can also be seen as an example of a singularly perturbed problem, as the type of the PDE changes from hyperbolic to hyperbolic-parabolic in the limit for large $k$. Note that the Burgers' equation should only serve as a limited example for the incompressible Navier-Stokes equations, in which the quadratic non-linearity in the convective fluxes is the most important term for the turbulent dynamics.\\
For the full Navier-Stokes system, a famous result obtained by balancing production and dissipation derived by Kolmogorov~\cite{kolmogorov1962refinement} states that the ratio of the smallest to the largest occurring scale is given by $\frac{\eta}{l_0}\propto Re^{-3/4}$. Taking this estimate to three spatial dimensions and accounting for the corresponding smallest temporal scales leads to the famous estimate for a lower bound for the number of degrees of freedom $\mathcal{N}$ required for a Direct Numerical Simulation (DNS) of free turbulence~\cite{rogallo1984numerical}: 
\begin{equation}\label{eq:re}
\mathcal{N}\propto n_{ppw}^4 Re^3,
\end{equation}
where $n_{ppw}$ is a measure of the accuracy of the discretization scheme. More important for our discussion here however is the scaling of the Reynolds number, which can make a full resolution of all occurring scales unbearably expensive. Thus, instead of solving the full governing equations, in the vast majority of practical applications, a coarse-scale formulation of the Navier-Stokes equations (NSE) is sought. These governing equations, i.e. the LES or RANS equations describing the coarse-scale dynamics and solution are found as follows: We start from a general form of a conservation law for a vector of conserved quantities $U(x,t)$, e.g. the Navier-Stokes equations, in the form
\begin{equation}\label{eq:les1}
R(U)=0.
\end{equation}
Here, $R()$ is the temporal and spatial PDE operator. We now introduce a low pass filter function $(\bar{.}) $ with linear filter kernel and a filter width $\bar{\Delta}$, such that we obtain a separation into coarse and small scales of a quantity $\phi$ as $\bar{\phi}$ and $\phi'=\phi-\bar{\phi}$, respectively. Applying this operation to Eq.~\ref{eq:les1} yields under some assumptions the coarse-scale governing equations for $\bar{U}$:
\begin{equation}\label{eq:les2}
R(\bar{U})=\underbrace{R(\bar{U})-\overline{R(U)}}_{=: M\approx\hat{M}(\bar{U},(\cdotp))}.
\end{equation}
Note that Eq.~\ref{eq:les2} incorporates both the RANS and LES methods, the difference being the choice of the filter. In RANS, a temporal averaging filter is chosen, returning the time-averaged or mean solution field. In LES, filtering is usually done in physical space, and a significant part of the temporal and spatial dynamics remain present. In both cases, the non-linearity of $U$ in $R$ prevents the commutation of filter and PDE operator and introduces the \emph{closure problem}: Although the LHS of Eq.~\ref{eq:les2} is now formulated in coarse scale quantities only, the RHS still contains a term that requires modeling:  $\overline{R(U)}$. This model term $\hat{M}$ as an approximation of $M$ must be formulated in coarse-scale quantities $\bar{U}$ (and possibly free parameters) only, otherwise the advantage of this approach would be lost. If of course $\overline{R(U)}$ was known exactly (e.g. from a precursor DNS simulation), then $M=\hat{M}$ would recover the exact closure term. This approach is termed "perfect LES" and is used as a consistent framework of model development~\cite{langford1999optimal}, but as it requires prior full scale information, it is not generalizable.  Note also that in order to solve for $\bar{U}$ numerically, $R$ must be replaced by its discrete equivalents. This entails a number of new difficulties, which will be discussed later. \\
Clearly, the introduction of the model $\hat{M}$ infuses uncertainties and errors into the solution process. Still, the rationale and justification for solving the coarse-scale equations instead of the original ones is given by
\begin{itemize}
	\item Universality of the modeled regime: The filtered out small scales are generally less affected by the boundary conditions and thus show a much more isotropic behavior than the large scales, which makes them amenable to modeling~\cite{batchelor1953theory}. In addition, the small scales carry only a fraction of the kinetic energy, while the large (resolved) scales are dominating the dynamics of the flow.
	\item Structural stability of the coarse scale solution: While single realizations of a turbulent field are highly sensitive to the initial and boundary conditions and small scale perturbations, the general structure of the solution is largely invariant to them. It can be shown that the truncated Navier-Stokes equations have attractors that share a number of properties to those of the continuous equations. This is the reason why the average and higher moments of the LES and DNS solutions are often in good agreement~\cite{temam1991approximation}.
	\item Computational necessity: Since Eq.~\ref{eq:les2} contains only coarse scale quantities (under the assumptions that the model regularizes the solution sufficiently and aliasing is accounted for), it is computationally considerably cheaper and avoids the requirement in Eq.~\ref{eq:re}. In fact, the computational costs can be specified a priori (within certain limits given by physical requirements) by choosing the filter width $\bar{\Delta}$.
\end{itemize}

\begin{figure}[htpb!]
	\centerline{\includegraphics[width=0.91\textwidth,trim= 0 0 0 0,clip]{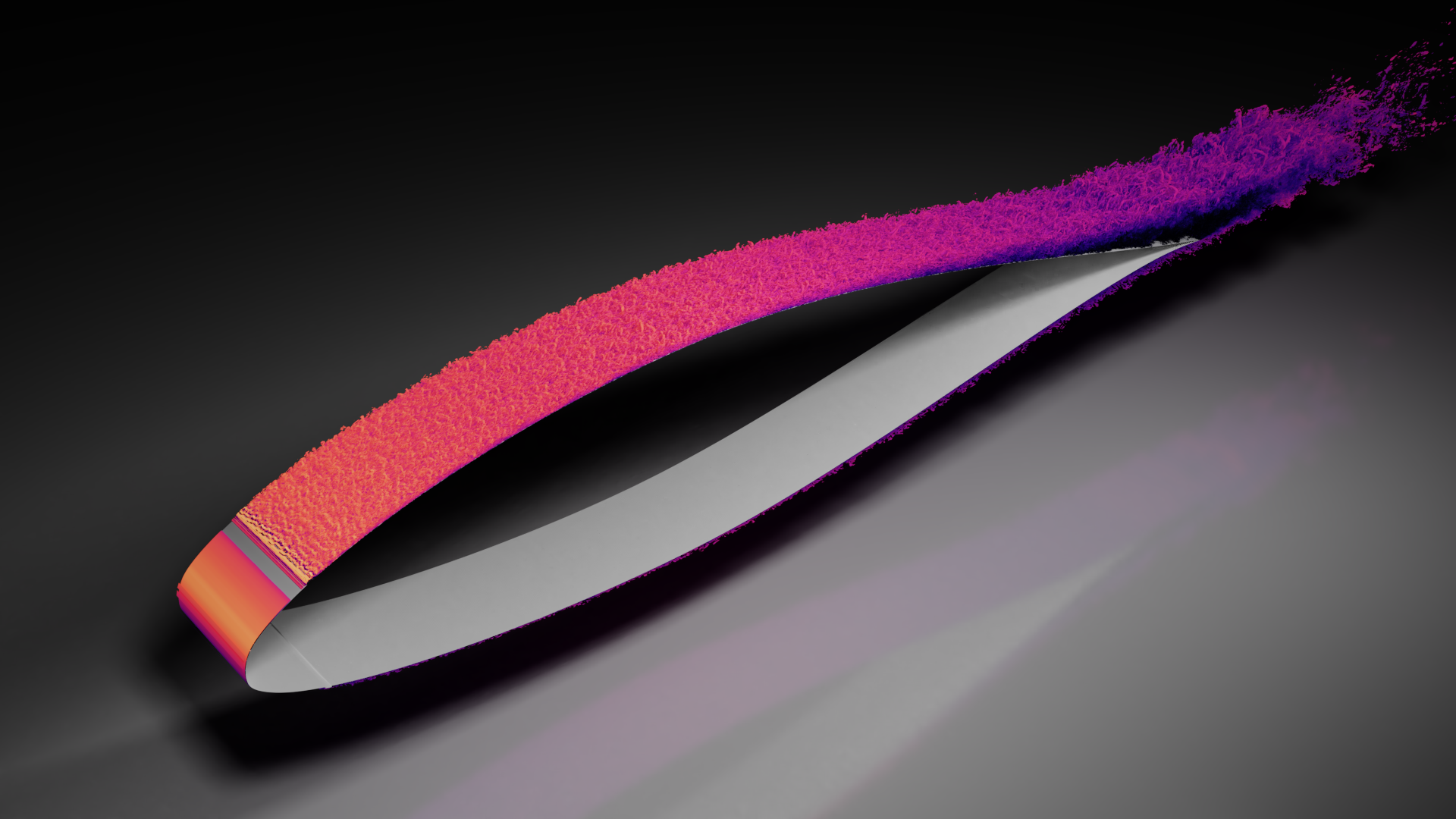}}\vspace{1mm}
	\centerline{\includegraphics[width=0.45\textwidth]{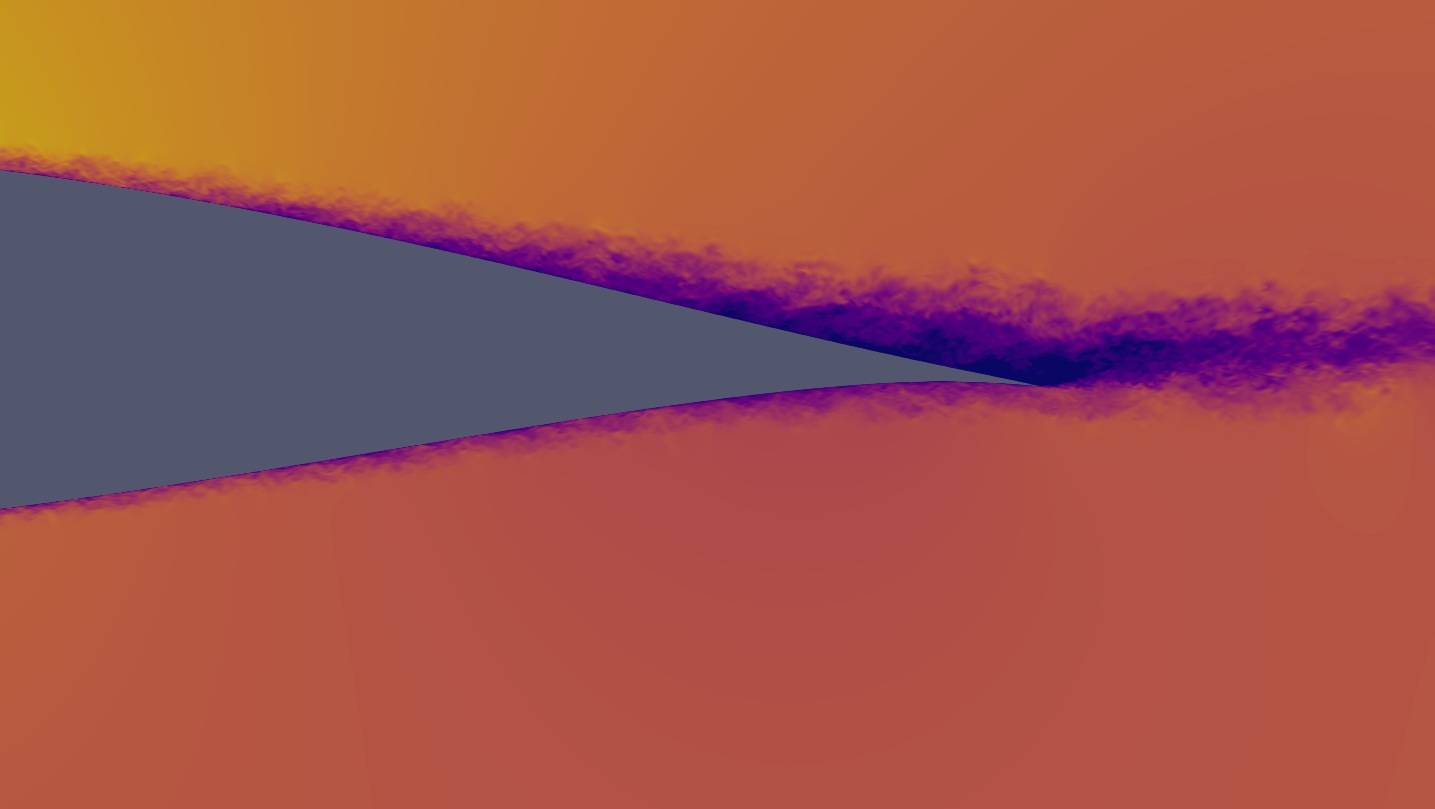}\hspace{1mm}\includegraphics[width=0.45\textwidth]{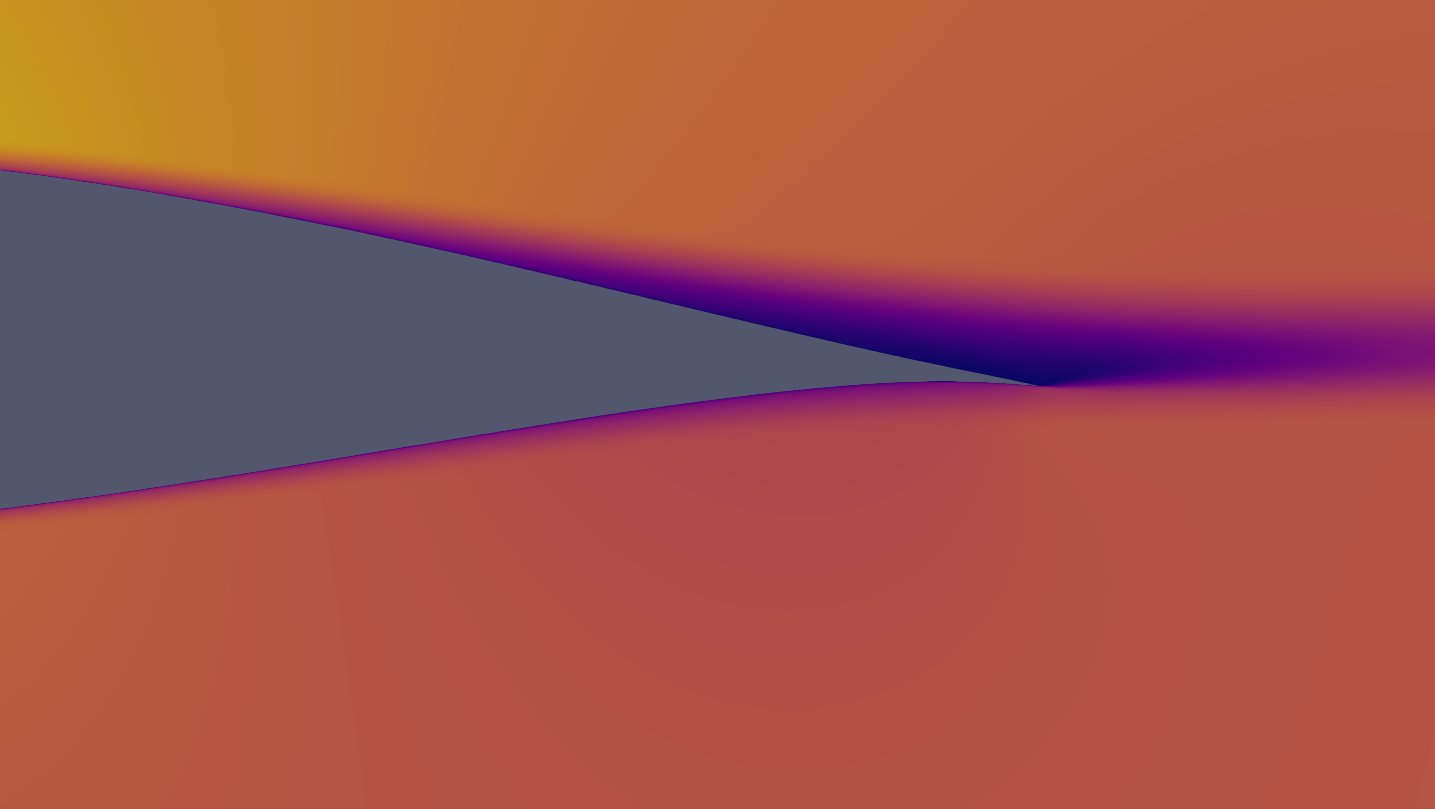}}
	\caption{Flow over a NACA 64418 airfoil at $Re_c=10^6$ and Mach number $Ma=0.2$. Top row: Underresolved DNS, shown are isocontours of the Q-vortex criterion, colored by velocity magnitude. Lower row: contours of vorticity magnitude, left: flow field from LES, right: time-averaged (RANS) results.\label{fig:airfoil}}
\end{figure}
\begin{figure}[htb!]
	\centering
	\includegraphics[width=0.98\textwidth]{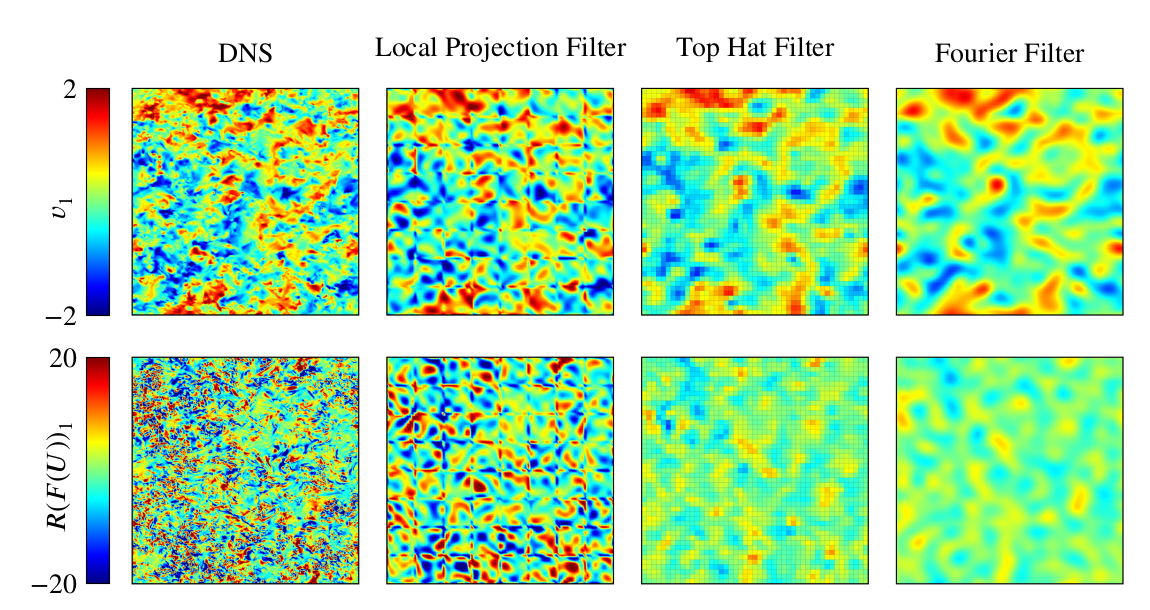}
	\caption{Two-dimensional slices of the full and the filtered x-velocity field~(\textit{top}) and $\overline{R(U)}$ flux divergence~(\textit{bottom}, $R(U)$ for DNS) from a DHIT flow. Shown is the field solution (from left to right) for the DNS, the local projection filter, the local top hat filter, and the global Fourier filter. Reproduced from~\cite{kurz2020machine}.}
	\label{fig:filter_field_solution}
\end{figure}

In order to visualize the model challenge and the RANS and LES solution approaches, Fig.~\ref{fig:airfoil} shows an example of the two different coarse-scale formulations applied to an airfoil flow at $Re=10^6$. In the top figure, an underresolved DNS, the multiscale character of the turbulent boundary layer is clearly visible. In the lower row, an LES (left) and a RANS (right) solution of the flow field in the vicinity of the trailing edge are shown. In both cases, the fine scale information is lost through the filtering process (all fluctuations are discarded in RANS) and both formulations require an additional model, however, the computational cost is reduced significantly as opposed to the full scale resolution.\\
While so far RANS and LES methods have shared the formalism, significant differences and challenges occur at closer inspection. First, while the filter in RANS is always defined as a temporal average and often the steady version of Eq.~\ref{eq:les2} is solved (we exclude the unsteady RANS approaches here), there is a lot more ambiguity in LES as one is essentially free to choose any meaningful filter with specific properties. One such example is the option to choose Favre-filtering for the compressible NSE. Another choice concerning the filter function directly are box filters, modal filters, discrete vs. continuous filters or even discretization-induced filters. A subtle but important point for model development is that any choice of filter always bring with it its own associated closure term $M$, which is clearly a function of the chosen filter~\cite{pruett2001toward}. Fig.~\ref{fig:filter_field_solution} underlines this property by showing results of the simulation of decaying homogeneous isotropic turbulence (DHIT), i.e. a canonical turbulent flow in a periodic box. The upper row contains the results for a velocity component $v_1\in \bar{U}$, filtered from the DNS result (left column) with three different filters (second to fourth column) typical for LES. The bottom row shows the associated contributions to the closure $\overline{R(U)}$ for the corresponding filter. The influence of the filter on the closure terms is clearly visible and stresses the role of the filter choice in LES. Additional complexities in the formal LES equations arise from a) the non-homogeneity of the filter (even if the filter kernel itself is shift-invariant, issues with grid stretching or at boundaries can occur) and b) the discretization operator and its associated errors. The action of the numerical schemes becomes the most important factor in so-called implicitly filtered LES. Here, the filter $(\bar{.}) $ is replaced by the discretization, which by design reduces the number of degrees of freedom of the full solution $U$ to a discretized version $\tilde{U}$. Note that while the $\tilde{(\cdot)}$ operator is often also called a filter (and the LES formulation is called "implicitly filtered" or "discretization/grid filtered"), its properties differ from $\bar{(\cdot)}$ in important aspects: Numerical errors occur, which are particularly active on the marginally resolved scales, and the discretization operators themselves can include non-linearities (e.g. for stabilization). This results in two issues worth noting: First, the actual solution $\tilde{U}$ is not known a priori and cannot be determined from DNS data through filtering (as is true for $\bar{U}$). Second, the choice of the discretization determines the closure terms (see also Fig.~\ref{fig:filter_field_solution}) and the discretization errors must become a part of their model. For a recent and more thorough discussion of these intricacies of LES, we refer the reader to~\cite{moser2020statistical,kurz2020machine}. Fortunately, the situation is less complex for the RANS method. Here, since only the mean solution is sought, the discretization effects and their interaction with the closure play a much lesser role, which allows the RANS modeling efforts to focus on "physics only", while for grid-filtered LES, both numerical and physical aspects must be considered. \\

Summarizing this discussion, the closure problem of turbulence can be described formally rather easily at first glance, but the resulting terms to be modeled can be influenced by a number of choices beyond the physical effects. In particular for grid-filtered LES, the numerical scheme, its errors and its induced scale filter contribute strongly to the exact closure terms. Thus, before attempting a data-based modeling approach, a careful appreciation and definition of the governing equations, their solution and the closure terms is necessary to ensure data-consistency.\\
Now returning to Eq.~\ref{eq:les2}, finding a model $\hat{M}\approx M$ is the next task. Here, in general, two approaches are common:
\begin{equation}\label{eq:model}
\hat{m_1}:=\min_{f(\theta),\theta} \left\lVert M-\hat{M}(\theta,f(\theta))\, \right\rVert\quad\quad \text{or}\quad\quad \hat{m_2}:=\min_{f(\theta),\theta} \left\lVert J(M)-J(\hat{M}(\theta,f(\theta)))\, \right\rVert
\end{equation}
In the first case, a best approximation to $M$ is sought directly as a function of some relationship $f(\theta)$ with parameters $\theta$, where both $f$ and $\theta$ can themselves be functions of the coarse scale fields. In the second case, the model is designed to approximate some derived quantity that is dependent on the model, e.g. an interscale energy flux or a spectrum. The most common form of models both for LES and RANS are from the $\hat{m_2}$ family and try to match the dissipation rate of the kinetic energy. Two major strategies exist for proposing the function $f$ in Eq.~\ref{eq:model}: it can be posited explicitly, i.e. some functional form based on physical or mathematical considerations is assumed (so-called explicit modeling), or the discretization error (possibly specifically modified) can replace $f$ (implicit modeling). In any case, due to the complexities in both the optimization process of Eq.~\ref{eq:model} and the interdependencies discussed above, the resulting models will be approximate and often only effective and accurate under circumstances which are comparable to the one present during their inception. A salient example of this is the fact that the constant for Smagorinsky's model has to be retuned for different discretizations and resolutions. \\
This section has served to highlight the challenges involved in model development and might help to explain why a converged state of the art in LES models is still lacking. In the next section, we give some ideas on why machine learning methods might help improve upon this situation.

\subsection{Turbulence modeling and machine learning}\label{subsec:why}
Before discussing the challenges and opportunities in ML and turbulence modeling, we give a brief and possibly incomplete working definition of ML for the purpose of this paper without going into the details of the different methods (cf. Sec.~\ref{sec:ml}):  

\begin{quote}
	Machine learning is a sub-field of artificial intelligence. It can be defined as a set of methods and algorithms for estimating the relationship between some inputs and outputs with the help of a number of trainable parameters. The learning of these parameters, i.e. their optimization with respect to a given metric, is achieved in an iterative manner by comparing the model predictions against groundtruth data or evaluation of the model performance. \end{quote}

From this definition, it becomes instantaneously clear that ML algorithms are \emph{not} the recommended method of choice if the relationship between inputs and outputs is a) known (e.g. given analytically) and easy to evaluate. In this case, ML methods typically suffer from two drawbacks: 1.) They are approximations to the analytical function only, even with possibly very high precision and 2.) they are not as efficient to evaluate as the analytical function. ML methods are also not recommended if b) the relationship between inputs and outputs is unknown but simple: For example, if a simple linear regression or quadratic curve fit will give sufficiently accurate results, then advanced ML algorithms will work but are unnecessary.\footnote{Linear regression can of course be formulated as an ML problem and a least squares fit is in fact the smallest building block of a neural network. However, in this context, we consider ML methods to include only more complex approximators like multilayered neural networks or support vector machines.}  Lastly and trivially, ML methods are not useful if c) input and output are not correlated. However, this statement must be accepted with some caution, even if there is only a negligible correlation between feature A and target Y as well as feature B and target Y, a feature made up of a non-linear combination of A and B may result in a high correlation to Y. In fact, this \emph{feature selection capability} is a particular strength of some of the ML methods.\\
In the case of the turbulence modeling challenge outlined above, neither of these hindrances are true, however. There is no closed known analytical formulation (on coarse grid data only) to close the equations, simple models work acceptably well but can be made to fail easily and we also know that coarse grid quantities correlate with the subgrid terms (in fact, the test filtering in the dynamic procedure of the Smagorinsky model is predicated on this). Therefore, it is at least reasonable to expect that ML methods are viable candidates to help in turbulence model building.  \\
Thus, a general application of an ML method to turbulence modeling can be formulated by specifying that the functional form $f$ and/or its parameters $\theta$ in Eq.~\ref{eq:model} are to be found by an ML method through training on data. As will also be discussed later, the functions that can be approximated by ML methods are quite rich, and finding an optimal function space that best represents the data is one of their strong points - thus, it is important to stress that ML methods cannot just fit parameters to some given basis, but also find the basis itself from data. This is somewhat orthogonal to the previously described modeling concept of explicit and implicit closure modeling in that no a priori functional form of the closure needs to be postulated. The training process of the ML model is then the optimizing w.r.t a given error norm based on the data available during the (offline) training phase. The application of an optimized model in a predictive manner (often called inference or model deployment) is then just an evaluation on a new data point. If the model is able to generalize well, it will give useful predictions for this unseen data. Here, careful observation of model-data-consistency is a key factor: the data during training (which implicitly defines the model) and inference must remain consistent. \\
With this brief conceptual overview of ML and turbulence modeling, we will now discuss some of the challenges and possible drawbacks but also chances of ML methods in turbulence. These lists are far from complete and are meant to raise general issues that may or may not be applicable to the specific case considered. It is expected that the surging research efforts in this field will help to answer these questions in the next years.

\subsubsection{Challenges for ML-augmented turbulence modeling}
In this section, we try to summarize the challenges and difficulties ML-augmented turbulence models must tackle. Not all of the issues are particular to ML, but casting them in an ML-context is useful for the purpose of this discussion.
\begin{description}
	\item[The need for data:] ML methods are at their core optimization algorithms that find an optimal fit of functions to data. What makes them so expressive is that the functions typically are high-dimensional and non-linear. This makes the optimization within the feature space costly, as the typical algorithms employed here are variants of a gradient-based, iterative progression to the minimum. The rate of convergence is thus limited and a function of the step size (called learning rate in ML), and there is no guarantee a global minimum has been found. Often in practice, several optimizations are run parallel to check if the found minimum is stable w.r.t. the hyperparameters and starting points. Thus, training an ML method requires significant amounts of data or a cheap repetition of the data generation mechanism. As both experiments and DNS simulations, which are the two "data generation" choices for turbulence, are rather costly, this poses a significant challenge\footnote{We note that there are attempts to synthesize turbulent fields through generative models, e.g. in~\cite{mohan2019compressed,PhysRevFluids.4.064603}, which potentially ameliorates this situation.}. Mining data from existing databases is also not straightforward due to the issues of data-consistency and data-model-consistency. In addition to the availability and  suitability of data, the question of cost effectiveness comes into play: It must be assured that the generation of training data constitutes a responsible use of computing time and experimental resources, and that the outcome (e.g. an improved turbulence model) justifies the investment. 
	\item[Consistent data and model:] Defining target quantities and input data for an ML method to learn from is not a straight forward task. As discussed above, while the solution and associated filter kernel to a RANS formulation is rather clear, the situation for practical, discretization-filtered LES is a lot more ambiguous. This also translates to any derived LES quantity computed from the solution field or through the application of the implicit filtering. This makes both the target of an ML method as well as its inputs dependent on the chosen LES formulation and discretization. In addition, the ML models must be made robust against training-inference inconsistency: Even if great care is taken to apply a model in a setting that corresponds to the one it was trained in, noisy data, the probabilistic parts of the ML method, error accumulation and the non-linear nature of the underlying turbulent process can lead to an input distribution that is vastly different from the one experienced in training. Even if the model predictions themselves are stable against disturbed inputs, their overall effect can be troublesome. For example, if the ML model predicts parameters in a closure model and the predictions themselves are reasonable and within range, the term the parameters are applied to can diverge from the training situation (e.g. a velocity gradient from training (DNS data) and become vastly different from the one during prediction (RANS/LES solution)). Another example is the influence of discretization operators discussed above. Thus, in general for an ML model to be successful, training and inference data must be consistent, the inference problem must be well-posed and the ML models must be made robust against unavoidable uncertainties. This is likely a crucial point in determining whether the advantages of ML-augmented models can be brought to bear.
	\item[Inclusive optimization:] Along the same vein, closure models must be considered on the level of the system of equations to be solved, not on a more fine-grained level. While this is true for any modeling approach besides ML-based attempts, it is worth pointing out here. This means that the overall closure must be optimized (also in the context of the discretization, see above), and that just improving the prediction of certain effects or terms does not necessarily lead to better models. It is instead the interplay of the different terms that must be considered. One example of this are the dynamics in turbulent boundary layers, where models that respect the overall balance of the physical effects have been shown to be superior to approaches that prefer the modeling of single contributions~\cite{larsson2016large}.
	\item[Physical constraints:] Incorporating physical and mathematical prior information and constraints into the models is likely important, as it not only helps with data-model-consistency or even consistency of the model to the governing PDEs, but also makes the resulting models more robust and accurate and easier to train. In addition, this information can help to reduce the amount of training data significantly by confining the parameter space to a sensible subspace. There are a number of ways to enforce constraints  in the ML process: The most obvious one is through the selection of training samples that share a given property, for example periodicity or positivity. This informs the ML methods of their existence implicitly, and helps to construct surrogates that also obey them - however, this is not guaranteed during inference or for extrapolated inputs. An additional measure is to explicitly include the required constraint like symmetry or realizability into the loss function and optimize accordingly. This usually leads to increased stability of the model, but again does not enforce the fulfillment of the constraint. So while constraining the input or the model output is the current state of the art, there is research underway to develop ML approximations with guaranteed properties, e.g.~\cite{santin2019kernel}. However, the relative importance of the respective constraints and the best way to enforce them remains an open challenge - not just for ML-based methods. For more information and a list of invariants and constraints, we refer the reader to~\cite{speziale_1985,oberlack1997invariant}.
	\item [Generalizability, interpretabillity and convergence:] As with all models, some basic questions must be addressed to ML-based formulations and their properties must be understood. Among these properties, generalizability is certainly one of the most desirable. This not only includes the applicability of a respective model outside of its training regime (i.e. at a Reynolds number not trained on), but also a consistent fallback mechanism in cases where the model is likely to fail. Even before that, a means of measuring confidence in the model prediction should accompany any model - in its simplest form, this could be an estimate of the position of the input data in feature space and a comparison against the statistics gathered during training. However, it is still unclear if we can expect more from ML in this context than from classical turbulence models, which all have their points of failure and generalization issues. The simplest case of generalization capability of an ML-augmented model should be at both limits of modeling: it should turn off in laminar flow and at the DNS resolution. A related but subtler property, in particular in conjunction with a discretization, is the question of the convergence of the model prediction with increasing flow resolution. Interpretability is probably the most illusive property on the list, and one may argue that a correct model prediction is more important than an understandable one. However, as it is unlikely that ML-augmented models will be ultimately successful without infusing expert human knowledge, understanding the ML-decision making is a necessary condition. Some general work on explainable ML algorithms can be found e.g. in\cite{doshi2017towards,zhang2018interpretable}.
	\item [Algorithmic and hardware considerations:] ML methods, in particular neural networks, strive on GPUs or even more specialized hardware (e.g. tensor processing units, TPUs) and in their native software environment like the Tensorflow~\cite{tensorflow2015-whitepaper} or PyTorch~\cite{NEURIPS2019_9015} suites, with Python being the usual language in which the user interacts with the underlying computational kernels. Legacy LES and RANS codes are on the other hand often written in C or Fortran, mostly with a focus on CPU execution, although GPU or hybrid codes are on the rise. In any case, during a practical turbulence simulation enhanced by ML methods, both classes of algorithms have to run concurrently, either on the same hardware or on specialized systems between which communication takes place. How to find optimal hardware and software stacks and how to balance the load between the flow solver and the ML kernels is an open field of research. Even tackling this question is currently difficult, as the state of turbulence modeling and ML (or any combination of traditional CFD and ML) is so fluent. In fact, as discussed in Sec.~\ref{sec:turb}, the role of ML in this field is yet undefined, and with this comes the uncertainty of "how much" ML will be incorporated into CFD. If only lightweight algorithms with very few parameters run locally, the question of how to incorporate them into a flow solver is dramatically different from the case when the full flow field itself or model terms are to be predicted by the ML methods. In the first limit, ML can be incorporated as an additional feature into the codes, in the second limit new flow solvers need to be designed around the ML algorithms themselves. Finding the best compromise here must be the next step after the possibilities and limits of ML for this application have been fully explored. 
	\item [Efficiency and ease of use:] Finally, from a practitioners point of view, the question that will likely determine the fate of ML-based models is: Is it worth it? A general observation is that not always the best model or approach finds widespread acceptance in the community, but rather the ones that are quite simple to understand and implement, robust to handle and computationally cheap. What might also make these models harder to adopt for non-specialists is the fact that they are not describable in a few lines of algebra or code, but instead as a computational graph and its parameters. This is of course a technical issue that can be solved, but it makes experimenting with the models much more cumbersome. So while ML methods have shown to be rather mighty in terms of accuracy, the other aspects are open for debate.  
	\end{description}

\subsubsection{Opportunities for ML-augmented turbulence modeling}
With the challenges identified in the previous section, we now aim to balance the discussion and give a perspective on the motivation and opportunities for ML-augmented turbulence models. In Sec.~\ref{sec:turb}, we will also present some successful data-based turbulence models and describe their relationship to the challenges and opportunities.
\begin{description}
		\item [A new paradigm:] Although considerable efforts have been invested in the last decades into finding closure models that are universal and accurate, there is still no generally accepted \emph{best} model. This is due to the complexities and inter-dependencies outlined above. All previous attempts at modeling based on proposing a functional form or a set of equations have shown that some form of data-assistance is necessary to make the model useful, typically through the introduction of parameters or the imposition of constraints (e.g. limiters). This reveals that there is still some form of functional relationship or knowledge not incorporated in these models - some external information seems to be missing. Here again, ML methods can help by attacking the problem from an alternate direction: The external data is not an afterthought to fix parameters, but an integral part of the model development itself. This is emphatically not meant as a replacement of mathematical theory or physical reasoning in the modeling process, but rather as an inclusion of an additional stream of information. 
	
	\item [Feature extraction:] While abstractions like the energy cascade, the universality of small scales, scaling behaviors or law-of-the-wall encapsulate the essence of the underlying mechanisms of turbulence and elucidate their effects, they can give a false sense of providing the full picture. Typically, physically motivated closure models attempt to exploit these known (or assumed) correlations. The most prominent example of this are eddy-viscosity based closures: Since the effects of small scale fluctuations are diffusive \emph{on average}, a model that shares this behavior \emph{on average} is a good initial guess. Following Boussinesq, the subfilter terms are then modeled as proportional to the mean velocity gradients, and model improvement comes from tweaking the proportionality constant. Machine learning can improve upon this situation not only be finding even better model constants, but more importantly, by identifying new correlations automatically (\emph{feature extraction}) from data, in particular very high-dimensional data. Thus, we can expect not only better tuned existing models, but also the discovery of a better functional basis for model building. 
	\item [Flexibility:] ML methods can be designed to deal with non-stationary and non-homogeneous statistics, i.e. they can approximate regime changes in space and time. This is highly interesting for turbulence, as this could help develop models capable of predicting intermittency and extreme events more reliably. 
	\item [Incorporating discretization effects:] An underappreciated aspect of LES is that a good closure model is an ever moving target. This is due to the fact that with very few exceptions, LES for practical problems is based on an implicitly filtered formulation. Here, the discretization itself acts as a low pass filter, and thus defines its very specific (and typically inhomogeneous and anisotropic) closure terms. Thus, a well-tuned and investigated closure approach for a given discretization can fail significantly in another setting. This strong non-linear interaction of model and numerical scheme adds additional dimensions to the closure problem - in a sense, this can be seen as the introduction of an additional non-linearity to the problem. While this makes the theoretical approach to finding models exponentially more difficult, it should - at least conceptually - be naturally included in an ML approach. 
	\item [Exploiting existing turbulence data:] As mentioned above, machine learning algorithms are not computationally cheap. In fact, one may argue that data-fitting approaches or learning methods are per se data-hungry during training - at least if the typical incremental learning processes are used. Although generating the training data from scratch through experiments or numerical simulation is possible, it is also prohibitively costly - after all, if the costs of providing enough samples to lead to a converged ML model are orders of magnitude larger than using established no-ML models in the first place, there is no justification for going the ML route. However, generating new data specifically for model training is likely unnecessary in the long run: A large treasure trove of flow data already exists, both from numerical and experimental sources, for example in~\cite{li2008public}. Mining this data, e.g. cleaning and processing it in a consistent manner before injecting it in the training cycle remains a challenge however. Here, the flexibility of ML approaches towards the input features and the inherent capability of non-linear feature combination can be beneficial in reducing the requirements on the training data.
	\item [Arbitrary input features:] ML methods are able to incorporate both temporal and spatial data naturally, and can also deal with sparse data. Up to now, typical closure models are often exclusively based on exploitation of spatial relationships or spatial data - for example a velocity gradient at a certain position and given point in time. This comes from the fact that a) the scale production mechanism that is at the root of the closure problem is formulated as a divergence and describes a spatial transport process and b) that the vast majority of scale-separation filtering in LES is formulated in physical space as well, with a notable exception being the work by Pruett et al.~\cite{oberle2020temporal}. Incorporating both dimensions into the modeling process is an attractive idea, as turbulence is a non-local phenomenon.

\end{description}
Summarizing this discussion, ML-based methods are attractive tools that can help to extract more, often hidden, knowledge from data, and thus provide a greater flexibility in modeling.  The promise of ML here lies in the incorporation of flexible, high-dimensional data and its natural use of non-linear modeling assumptions. Still, a number of challenges exist that need to be tackled, as essentially in any modeling process, and no converged state of the art has been reached. Therefor, we hope that these lists provided here can help guide model development and evaluation in the future.\\
After briefly introducing selected ML algorithms for context in the next section, we will then discuss some successful applications of ML methods to turbulence modeling in Sec.~\ref{sec:turb}. This should not only represent a slice through the current state of the art, but also show the diversity of the ideas and methods currently under investigation.

\section{Machine Learning Algorithms}\label{sec:ml}
The terms \textit{Artificial Intelligence} (AI), \textit{Machine Learning} (ML) and \textit{Deep Learning} (DL) are often used synonymously, due to the tremendous success of the latter two in the last decades, making clear distinctions and definitions beyond general statements like the one given in Sec.~\ref{subsec:why} elusive. In principle, ML is only one specific discipline of many in the broad realm of AI, but of course currently the most prominent one.
A complementary field in AI is e.g. ``Good Old Fashioned AI'' (GOFAI), which encompasses AI systems that can be written down in a symbolic, human-readable way\cite{Haugeland1990}.
They therefore build on explicit domain expertise incorporated into the systems.
One of the most famous representatives of this field are the \textit{expert systems}.
Another famous example for AI without the use of ML is the DeepBlue system, which was able to beat at the time chess world champion Garry Kasparov in 1997 by using AI~\cite{Campbell2002}, but without a hint of ML.\\
In contrast, ML algorithms are (at least to a certain degree) agnostic to the specific task they are used for.
Rather, they are able to learn from experience and adapt themselves to (or \textit{learn}) the specific task autonomously based on data.
This is of course an idealized view, since choosing a better suited ML algorithm over another for a specific task based on experience already incorporates some form of expertise.
The ML approach obviously has some advantages:
foremost, machine learning algorithms can be employed where no sufficient domain knowledge is available to build expert systems.
Additionally, existing algorithms can be used for a variety of tasks that are similar in structure (e.g. using a neural network that is trained on detecting visual edges in photographs can very easily be brought to recognize shock waves as sharp edges from flow field data~\cite{shock}) and only have to learn to solve the new problem, while other AI algorithms developed the traditional way  might be useless for the new task.\\
The field of ML can be categorized according to several learning paradigms: 
The first is supervised learning (SL), which is discussed in Sec.~\ref{sec:SL}.
The foundation of an SL task is always an assumed correlation between at least two quantities, an input and an output.
It is then the ML algorithm's task to approximate this unknown, but probably present functional connection between these quantities, solely from sampled training data.
In contrast, unsupervised learning (UL) tasks, which are detailed in Sec.~\ref{sec:UL}, do not need labeled data, i.e. data where the exact output is known. 
This learning paradigm is of more exploratory nature and confines itself to identifying correlations and patterns inside data, as well as finding efficient low-dimensional representations.
The last learning paradigm is reinforcement learning (RL), which does not rely on a dataset, but rather employs algorithms which learn by interacting with an environment.
The RL concept is discussed further in Sec.~\ref{sec:RL}.\\
We are aware that the borders of these categories are fluent and the assignment of applications and algorithms to the specific fields is more often than not ambiguous.
Moreover, many methods do not fit comfortably into this framework but rather establish a small sub-category on their own.
Nonetheless, this segmentation follows the usual classification and gives orientation for more in-depth excursions in ML.
In the following discussion, we put a special emphasis on supervised learning with neural networks, since they have become the de facto state of the art for many applications, among them turbulence modeling.
\subsection{Supervised learning}
\label{sec:SL}
\begin{figure}[t]
	\centerline{
    \includegraphics[width=0.75\textwidth]{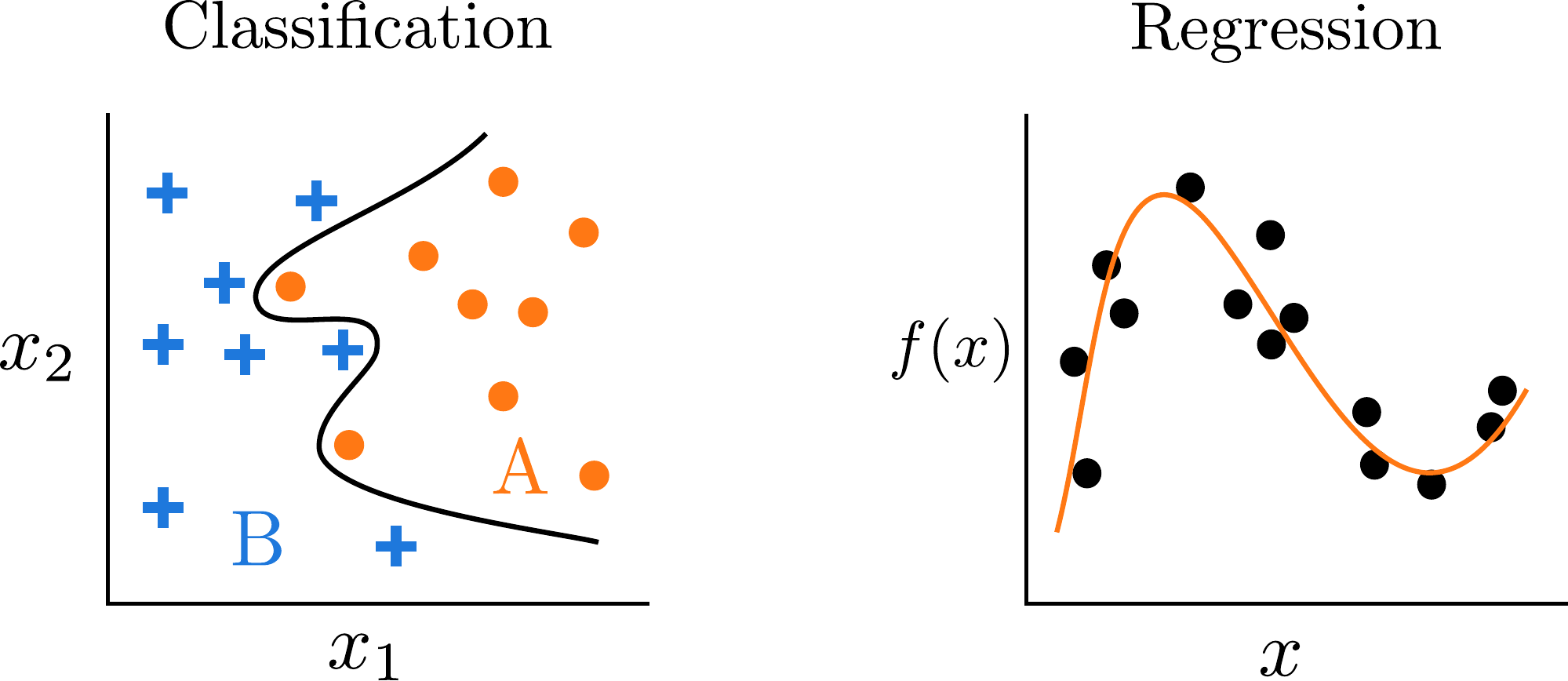}
  }
  \caption{The two major paradigms of supervised learning: regression und classification. For a classification task, the ML algorithm needs to find a decision boundary, which separates the data into classes. New data points can then be classified by determining on which side they reside. For regression, the algorithm approximates the continuous functional relationship of the data.}
  \label{fig:supervised_learning}
\end{figure}

The basis of a supervised learning task is a dataset of samples $\left\{X_i,Y_i\right\}$ from some unknown functional input-output relationship $Y=f(X)$.
The task of the supervised ML algorithm is to approximate the true function $f(X)$ solely based on this (possibly noisy) dataset, which is commonly called training data.
For continuous data this task is called regression.
Once a good approximation to the true function is found, the model can be evaluated for data points, which were originally not part of the training set (the \emph{generalization}).
If the data is discrete, i.e. the algorithms has to assign the input data to a finite amount of classes, this task is referred to as a classification task.
The objective of the algorithm is then to find some (non-linear) decision boundary in the input space, which separates the data points of the different classes from each other.
Based on this decision boundary, new unknown data can be classified by determining on which side of the decision boundary they reside.
Both approaches of supervised learning are shown in Fig.~\ref{fig:supervised_learning}.
\\
As discussed above, ML algorithms are agnostic to the underlying learning task and can be applied to a wide variety of problems.
To accomplish this, ML models comprise free parameters, which have to be adapted for each learning task.
The search of good model parameters for a specific learning task can be formulated as an optimization task with regard to some performance metric called \textit{loss function} $\mathcal{L}(\hat{Y},Y)$ or $\left\lVert Y-\hat{Y} \right\rVert$, see Eq.~\ref{eq:model}.
A common loss function for a regressor is e.g. the $L_2$-distance between $Y$ and $\hat{Y}$.
The optimization task can therefore be interpreted as finding a set of model parameters $\theta$, for which the loss function reaches a minimum on the given training set $\left\{X_i,Y_i\right\}$:
\begin{equation}
  \theta_{opt} = \underset{\theta}{\mathrm{argmin}} \:\mathcal{L}\left(\:\theta \:\left| \left\{X_i,Y_i\right\}\right.\right) \:.
\end{equation}
In the context of ML, this optimization process is commonly referred to as \textit{learning} or \textit{training}.
A supervised ML algorithm therefore has two distinct phases: in the learning phase, the free model parameters are adapted to minimize the loss function for the given training dataset. In the inference phase, the trained model is evaluated on \emph{unseen} data to obtain predictions for the corresponding unknown output value.
In the following, we will shortly introduce some of the most common ML methods for supervised learning: Random forests, support vector machines and artificial neural networks, with special emphasize on the latter.
\subsubsection{Random forests}
\begin{figure}[t]
	\centerline{\includegraphics[width=0.75\textwidth]{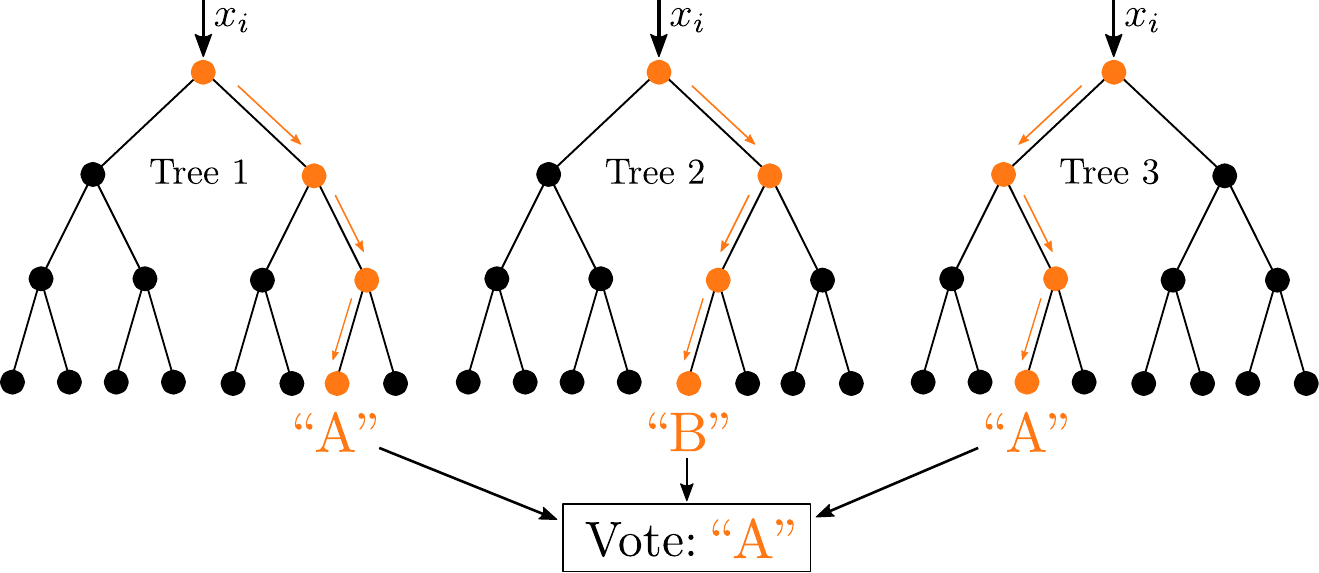}}
	\caption{A schematic view of the random forest method. The depicted random forest model comprises three independently trained decision trees, which receive the same unseen input data $x_i$. The infered class of $x_i$ is then derived from the individual predictions by a majority vote.}
  \label{fig:random_forests}
\end{figure}

Decision trees are one of the most intuitive machine learning algorithms.
This ML algorithm classifies data by dividing the training set recursively according to decision rules, until each of the resulting subsets contains only data of one distinct class.
Training the decision tree refers to finding a set of decision rules which divide the training set most efficiently.
New unseen data can then be classified with the same decision rules and the model can propose an assumed class for the respective input data.
Since it is not trivial to ensure that the found decision rules rely on intrinsic characteristics of the data and therefore to generalize well to unseen data, decision trees are prone to overfitting and are highly sensitive to changes in the training dataset.
Therefore, more elaborate variants of decision trees are proposed in literature, with \textit{random forests} as one of the most famous representatives.
The method of random forests was originally proposed by Ho\cite{Ho1995}, who extended the single decision tree to an ensemble of individual decision trees to improve the generalization properties of the method.
The main idea is that each individual decision tree is trained only on a random subset of the training data, or receives only a randomly selected subset of the data features as input data.
The overall class label for the data is then obtained by a majority vote of the individual predictions, which significantly enhances the generalization abilities of the decision tree method (cf. Fig.~\ref{fig:random_forests}).
The idea to combine individual models to ensembles in order to enhance the overall prediction accuracy can also be employed for other algorithms, e.g. neural networks.
The decision tree and random forest algorithms can be easily extended to regression tasks by replacing the majority vote by averaging the individual predictions.
In the field of turbulence simulation, random forests were e.g. applied by Wang~et~al.\cite{Wang2017}, who trained the model on DNS data in order to correct the Reynolds stresses of RANS simulations.
%
\subsubsection{Support vector machines}
\begin{figure}[t]
	\centerline{\includegraphics[width=0.75\textwidth]{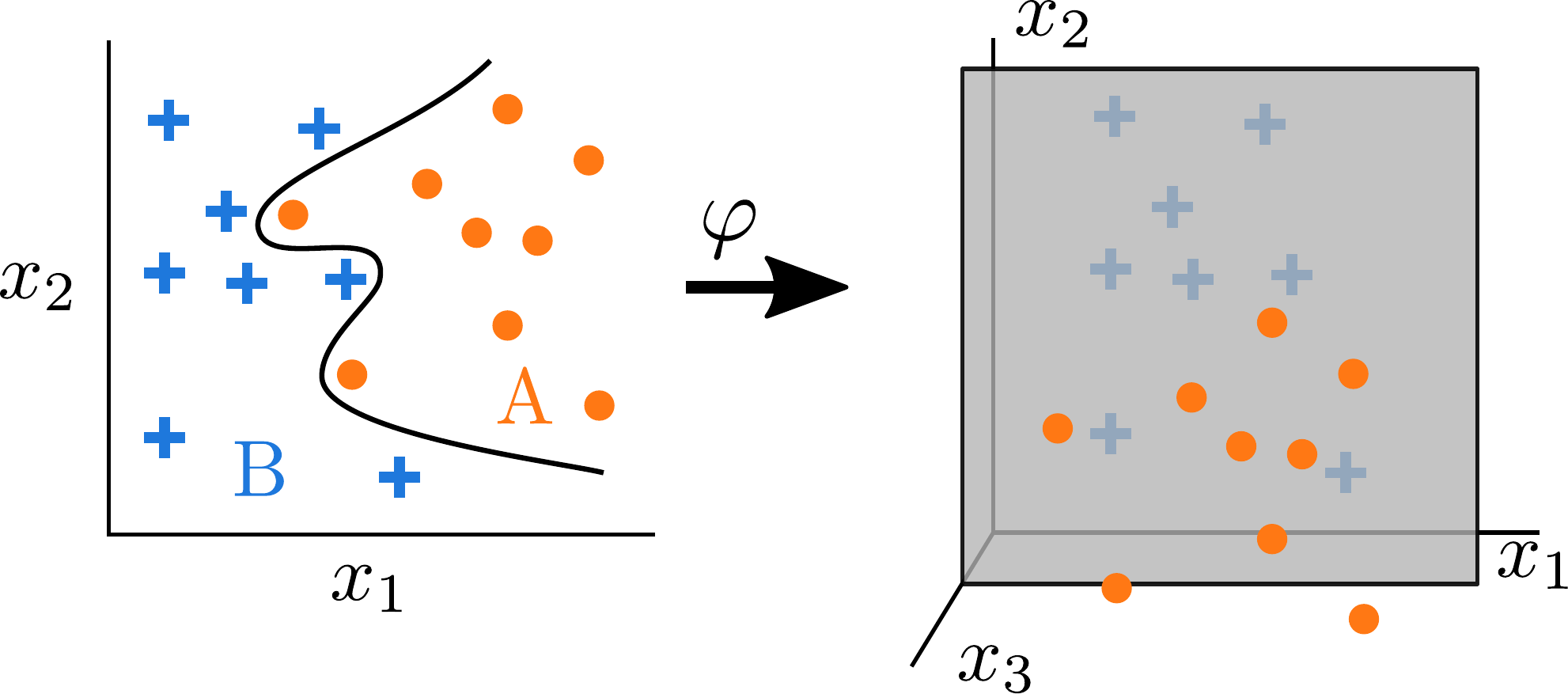}}
  \caption{General idea of a high-dimensional feature space. Some non-linear mapping $\varphi$ is used to transform the two-dimensional input space into a three-dimensional feature space, in which the data becomes linearly separable. The inverse mapping then transforms the linear plane into the non-linear decision boundary in the two-dimensional input space on the left.}
  \label{fig:kerneltrick}
\end{figure}

The support vector machine (SVM) algorithm strives to find a hyperplane that separates the space spanned by the input features such that only data points of one distinct class reside on each side of the hyperplane.
New unseen data can then be classified by determining on which side of the separating hyperplane the respective data point resides.
For most training datasets however, the classes are obviously not linearly separable.
Boser~et~el.\cite{Boser1992} therefore proposed to employ a non-linear mapping of the input space in a high-dimensional (and possibly infinite-dimensional) feature space, in which the training data will always be linearly separable by a hyperplane, as indicated in Fig.~\ref{fig:kerneltrick}.
These computations in high-dimensional space are generally expensive.
The so-called \textit{kernel trick} can be used to keep the computational cost at reasonable levels.
By choosing an appropriate kernel function, computations in the high-dimensional feature space can be carried out implicitly by applying the kernel function, while the coordinates of the data points in the high-dimensional feature space never have to be computed explicitly.
SVM were primarily applied for supervised classification tasks, but can also be transferred to regression\cite{Smola2004} and to unsupervised tasks, e.g. clustering\cite{Ben2001}.
One of their most interesting features is the capability to enforce constraints directly through the kernel choice\cite{Scholkopf1998}.
A more in-depth introduction to SVM can be found in\cite{Steinwart2008}.
In~\cite{Ling2015}, SVM are used together with decision trees and random forests to predict in which flow regions certain assumptions of RANS models become invalid, which can e.g. be used to dynamically adapt turbulence models based on the prevalent flow regime.
\subsubsection{Artificial neural networks}\label{subsec:ann}
\begin{figure}[t]
	\centerline{
    \includegraphics[width=0.34\textwidth]{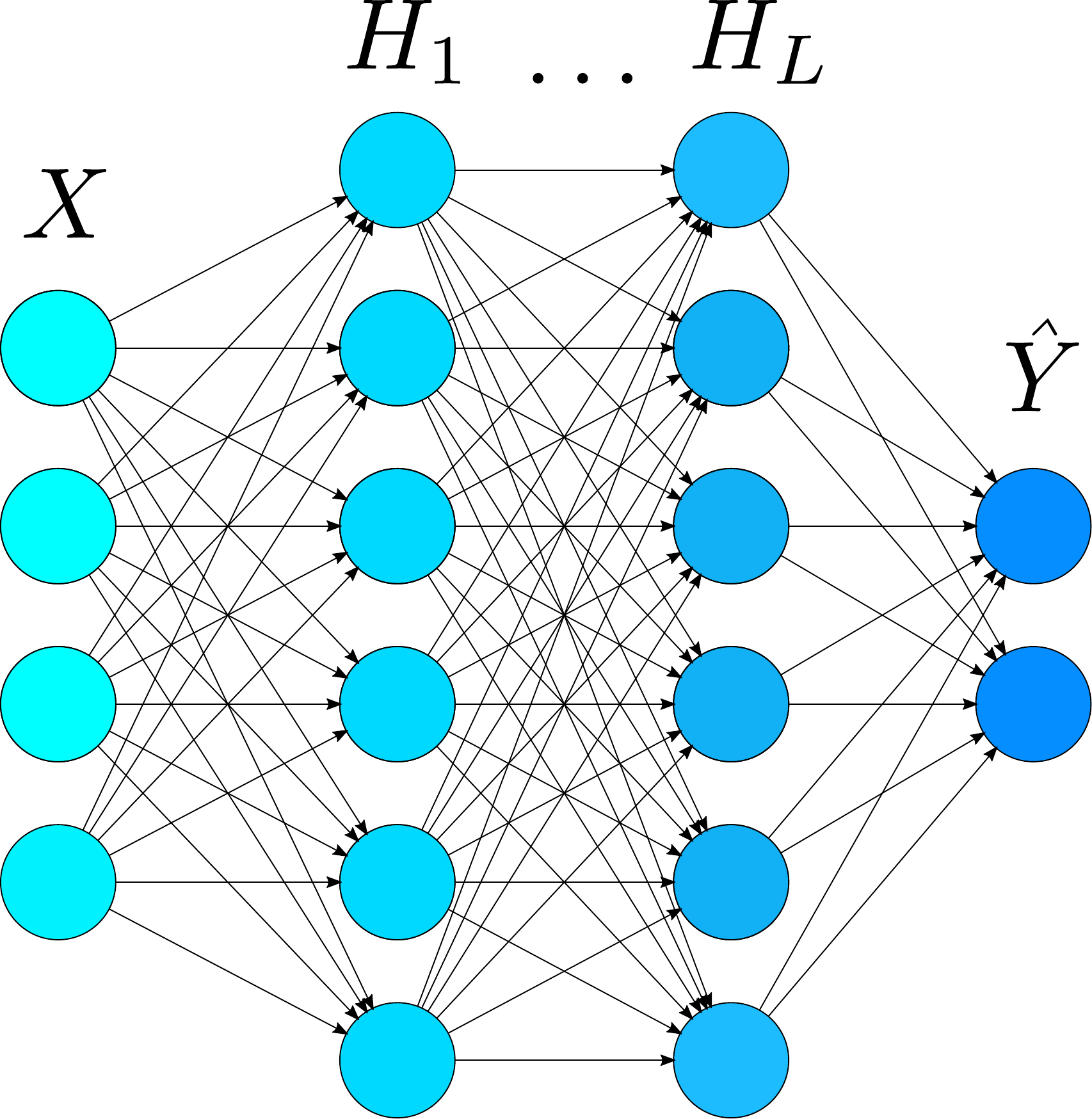}
    \hspace{10mm}
    \includegraphics[width=0.45\textwidth]{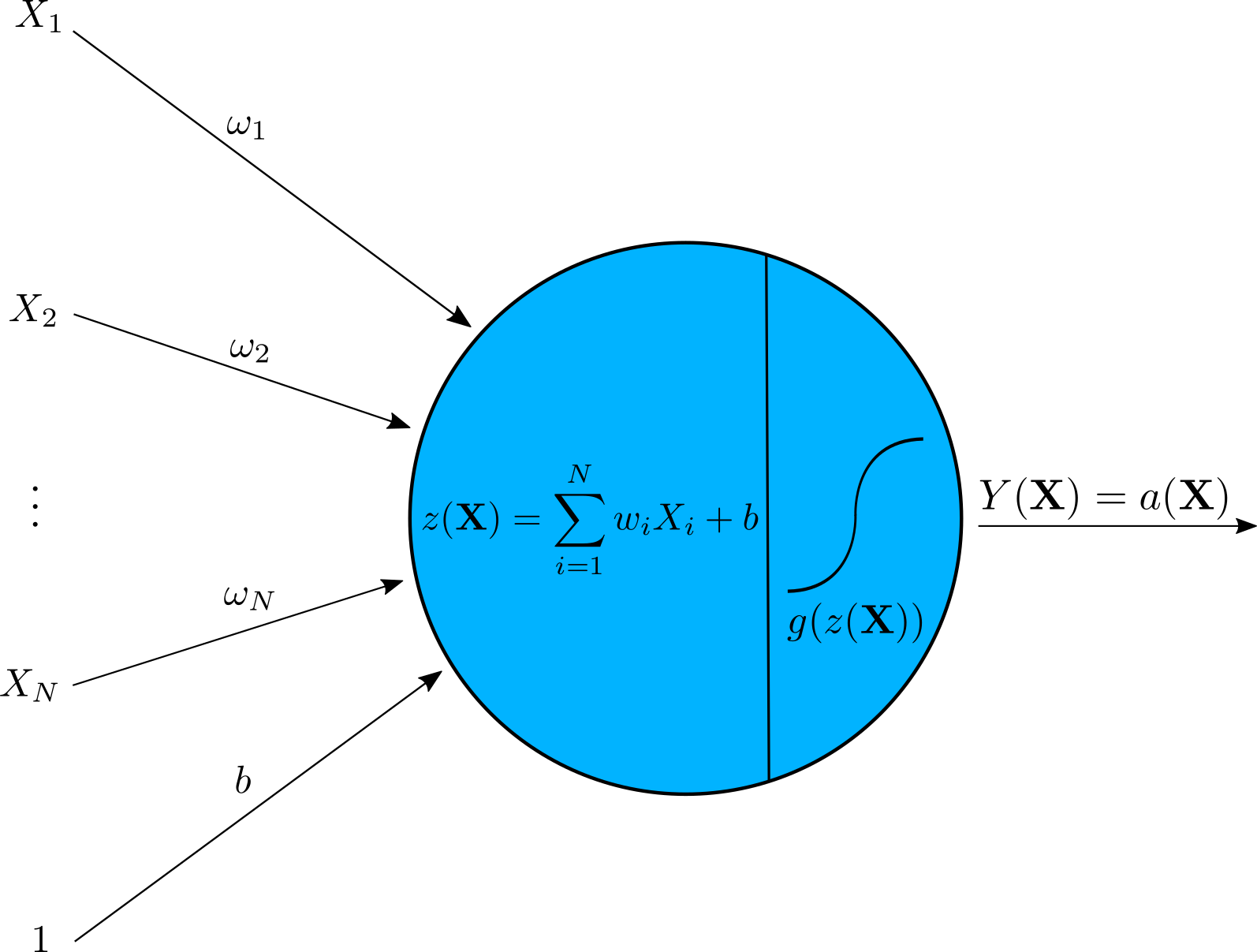}
  }
	\caption{Shown on the left is the layout of a general feed-forward ANN with two hidden layers, each comprised of several neurons. The design of a perceptron neuron is shown in more detail on the right.}
  \label{fig:mlp}
\end{figure}

\begin{figure}[t]
	\centerline{
    \includegraphics[width=0.55\textwidth,trim=0 0 100 0, clip]{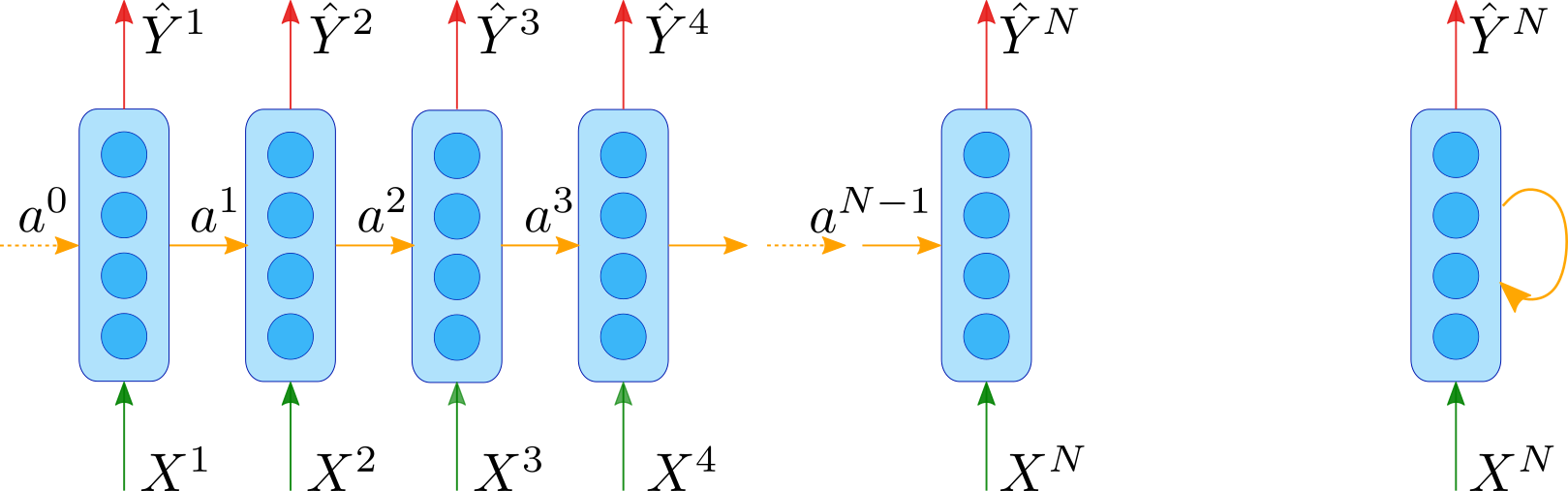}
    \hspace{6mm}
    \includegraphics[width=0.35\textwidth]{./figures/ml/LSTM_cell.pdf}
  }
  \caption{Shown on the left is the schematic of the RNN approach when unrolled in time. Along with the current sample of the sequence $X^N$, the RNN receives information from the last timestep $a^{N-1}$, which allows it to retain temporal correlations from the sequential data. For the LSTM cell on the right, the rectangular shapes indicate layer-wise operations according to Eq.~\eqref{eq:perceptron} with either a sigmoid ($\sigma$) or a hyperbolic tangent (tanh) as activation function. Round shapes imply element-wise operations, with $\mathbin\Vert$ indicating concatenation. For the LSTM, the retained information~$a^N$ consists of the cell state~$c^N$ and the hidden state~$h^N$, while the latter is also the layer output $\hat{Y}^N=h^N$.}
  \label{fig:RNN}
\end{figure}

Artificial neural networks (ANN) are the most widespread machine learning algorithm for many fields and applications, especially due to their generality and ease of training on GPUs.
Their name is derived from the somewhat constructed resemblance to structures in mammalian brains~\cite{Mcculloch1943,Rosenblatt1958}, which was the initial motivation for the specific design of ANN as a combination of individual "neurons", shown in Fig.~\ref{fig:mlp}.
The classic feed-forward network is comprised of several concatenated layers. Each layer in turn consists of a number of neurons.
The input layer of the network receives a data vector $X$, which is then successively passed through the \textit{hidden layers}, which are not directly connected to the input or output, until the data is passed to the output layer.
ANN with few hidden layers are referred to as \textit{shallow} ANN, while networks with many hidden layers are called \textit{deep} ANN.
A typical design for the neurons is the perceptron as shown in Fig.~\ref{fig:mlp}, which was originally formulated by Rosenblatt\cite{Rosenblatt1958} as simple mathematical model for neural activity.
Each perceptron receives the outputs of the previous layer $X_i$ as inputs.
These inputs are weighted by \textit{weights} $\omega_i$ and summed up with an additional bias $b$.
The output (also called activation) of the perceptron is then obtained by applying a non-linear \textit{activation function}.
Commonly used activation functions are sigmoidal functions like the sigmoid ($g(x)=1/\left(1+e^{-x}\right)$) or the hyperbolic tangent ($g(x)=\mathrm{tanh}\left(x\right)$).
More recently, the rectified linear unit (ReLU, $g(x)=\mathrm{max}\left\{x,0\right\}$) became the state-of-the-art for many applications.
The computed output of the perceptron is then
\begin{equation}
  Y = g\left(\sum_{i=1}^N \omega_i X_i + b \right).
  \label{eq:perceptron}
\end{equation}
The output $Y$ is passed as input for the neurons in the succeeding layer.
The main objective during training is to determine good weights and biases to obtain accurate predictions from the network. Thus, in short, a neural network is a nested sequence of linear and non-linear functions with variable parameters. In general, artificial neural networks are \emph{universal approximators} and can approximate any continuous functional relationship between input
and output quantities solely based on data and without prior assumptions on the nature
of said function~\cite{cybenko1989approximation,hornik1991approximation,lu2017expressive}.
\\
A particularly successful branch of ML emerged over the last decade: deep learning (DL).
The method DL exploits that deeper networks can learn more complex relationships, since each additional layer provides a (potentially) more abstract and complex representation of the data \cite{LeCun2015}.
Training such deep and parameter-rich networks was made possible by a variety of circumstances.
Firstly, the method of backpropagation yields a very efficient way of optimizing the model's weights.
The more recent success was also supported by the steadily increasing computational power, in particular by employing highly parallel graphic processing units (GPU) for training~\cite{Schmidhuber2015}.
The other major contributing factor was the availability of large datasets, which are needed to train deep networks, since they comprise an enormous amount of free parameters, and thus tend to overfit on small datasets\cite{LeCun2015}.
These advancements fueled the research and development in the field of deep learning and caused it to become the most prominent and successful representative of machine learning for many applications.\\
A number of specialized ANN architectures exist, which are designed for a specific range of tasks. For multi-dimensional data, e.g. images, convolutional neural networks (CNN) are the current state of the art \cite{LeCun2015,Schmidhuber2015}.
In contrast to the perceptron architecture, CNN apply filter kernels to the input data in each layer, while the filter kernels themselves are learned during training.
This generally results in a more sparse connection of succeeding layers and therefore in less trainable weights than in classic fully-connected networks.
This approach is especially suited for multi-dimensional data, which is not point-local, but rather has to be examined in context of its surrounding.
Very deep ANN have shown to be more difficult to train, since they suffer from the vanishing and exploding gradients problem during backpropagation, see e.g. \cite{Schmidhuber2015}.
To alleviate this restriction, He~et~al.\cite{He2016} proposed the residual neural network, which introduces skip connections, allowing information to shortcut the non-linear layers. 
The CNN layers therefore only have to approximate the non-linear fluctuations, while linear relationships are approximated quickly and stably. 
This allowed to design and train very deep CNN successfully, thereby allowing to build more expressive and accurate CNN\cite{He2016}.\\
For sequential data, i.e. where the ordering of the data points is important, recurrent neural networks (RNN) are the established ANN architecture.
A general schematic of the RNN principle is shown in Fig.~\ref{fig:RNN}.
The standard RNN works like the perceptron network shown in Fig.~\ref{fig:mlp} when unrolled in time.
However, along the current input sample of the sequence $X^N$, the RNN also receives some information $a^{N-1}$ from the last sample, which enables the RNN to memorize information from previous samples.
How the network output $Y^N$ and the memorized state $a^N$ are computed depends on the specific RNN type.
For long sequences, the RNN architecture leads to very deep networks in time, which again rises the issue of vanishing and exploding gradients during backpropagation, rendering the standard RNN unsuitable for long sequences~\cite{Hochreiter1997}.
More elaborate variants of RNN resolve this problem by introducing gating mechanisms which control the error transport during backpropagation.
Two of the most common variants are the gated recurrent units (GRU)\cite{Cho2014} and the long-short term memory (LSTM) network\cite{Hochreiter1997}, which is detailed in Fig.~\ref{fig:RNN}.
\subsection{Unsupervised learning}
\label{sec:UL}
\begin{figure}[t]
	\centerline{
	  \includegraphics[width=0.7\textwidth]{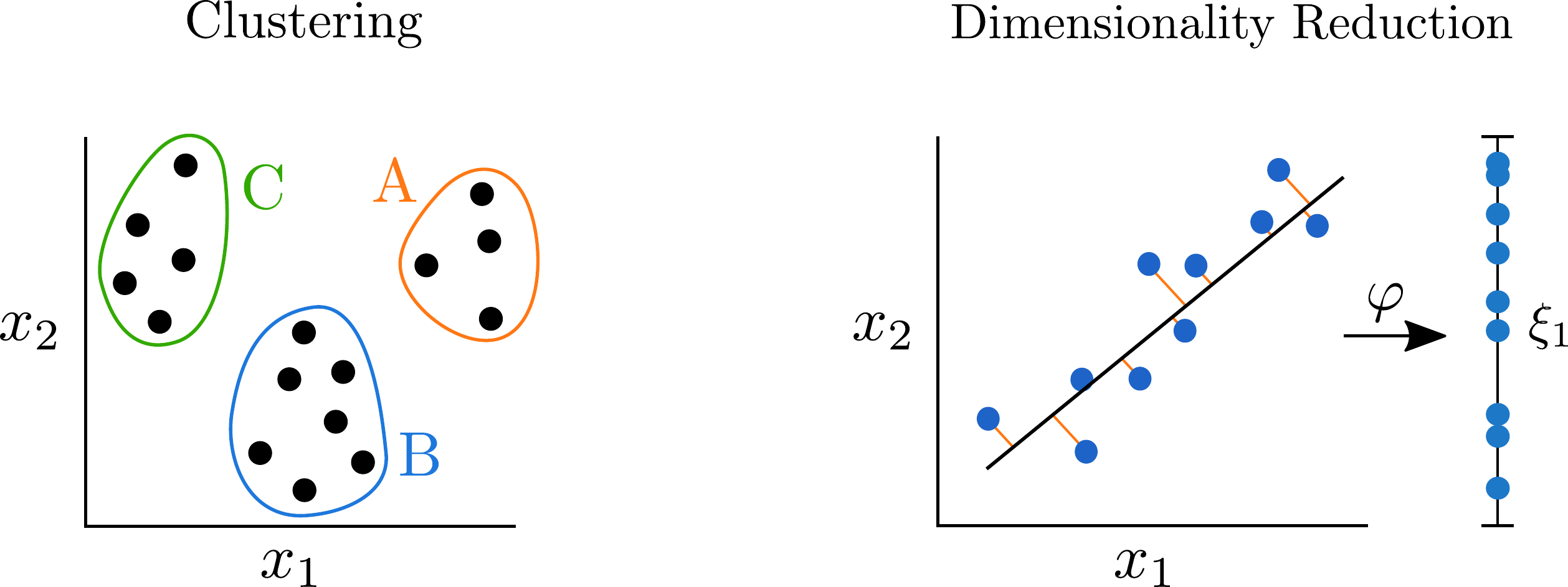}
  }
  \caption{Two main applications of unsupervised learning: clustering and dimensionality reduction. For clustering, points with high similarity (i.e. near each other) are grouped together in clusters to identify patterns in data. The field of dimensionality reduction tries to transfer high-dimensional data into efficient low-dimensional representations by finding appropriate transformations $\varphi$.}
  \label{fig:unsupervised_learning}
\end{figure}
In contrast to the SL paradigm, unsupervised learning (UL) does not need labeled data, i.e. it does generally not require any information on the exact solution of the problem.
A common area of application for UL is clustering, which is the task of grouping data points based on their similarity, which in consequence allows to find unknown correlations and patterns in datasets. 
In the following section, the k-means algorithm will be discussed exemplary as a common and straight-forward clustering algorithm. 
However, also clustering variants of several SL algorithms like SVM\cite{Ben2001} can be found in literature.
Another major task of UL is the field of dimensionality reduction, which targets at finding efficient low-dimensional representations of data, while minimizing the resulting information loss - this is of course closely related to finding a cluster of data with similar features.
Both concepts are illustrated in~\ref{fig:unsupervised_learning}.
Other applications of UL like probability density estimation are not reported here, since this exceeds the scope of this work.
\subsubsection{k-means}
A popular and fairly simple clustering algorithm is the \textit{k-means} method\cite{Macqueen1967}, for which a vast number of variations and extensions are proposed in literature.
The first step of the algorithm is to obtain a first guess of the $k$ partition centroids by e.g. choosing $k$ data points at random as centroids or by more elaborate approaches~\cite{Arthur2006}.
Secondly, all data points are assigned to the respectively nearest cluster centroid, and thus each data point is assigned to one of the $k$ clusters.
In a third step, the cluster centroids can be updated by computing the center of all points assigned to the respective cluster.
The second and third step are then repeated until the algorithm converges and the clusters cease to change.
Thus, the k-means algorithm eventually minimizes the average variance in the clusters.
More elaborate variants of the k-means algorithm increase the method's accuracy and allow to give guarantees and estimates of the method's performance, as e.g. k-means++\cite{Arthur2006}. An application of k-means clustering in conjunction with a NN for the prediction of an eddy-viscosity for the RANS equations can be found in~\cite{zhu2019machine}.
\subsubsection{Autoencoder}
\begin{figure}[t]
	\centerline{
	  \includegraphics[width=0.5\textwidth]{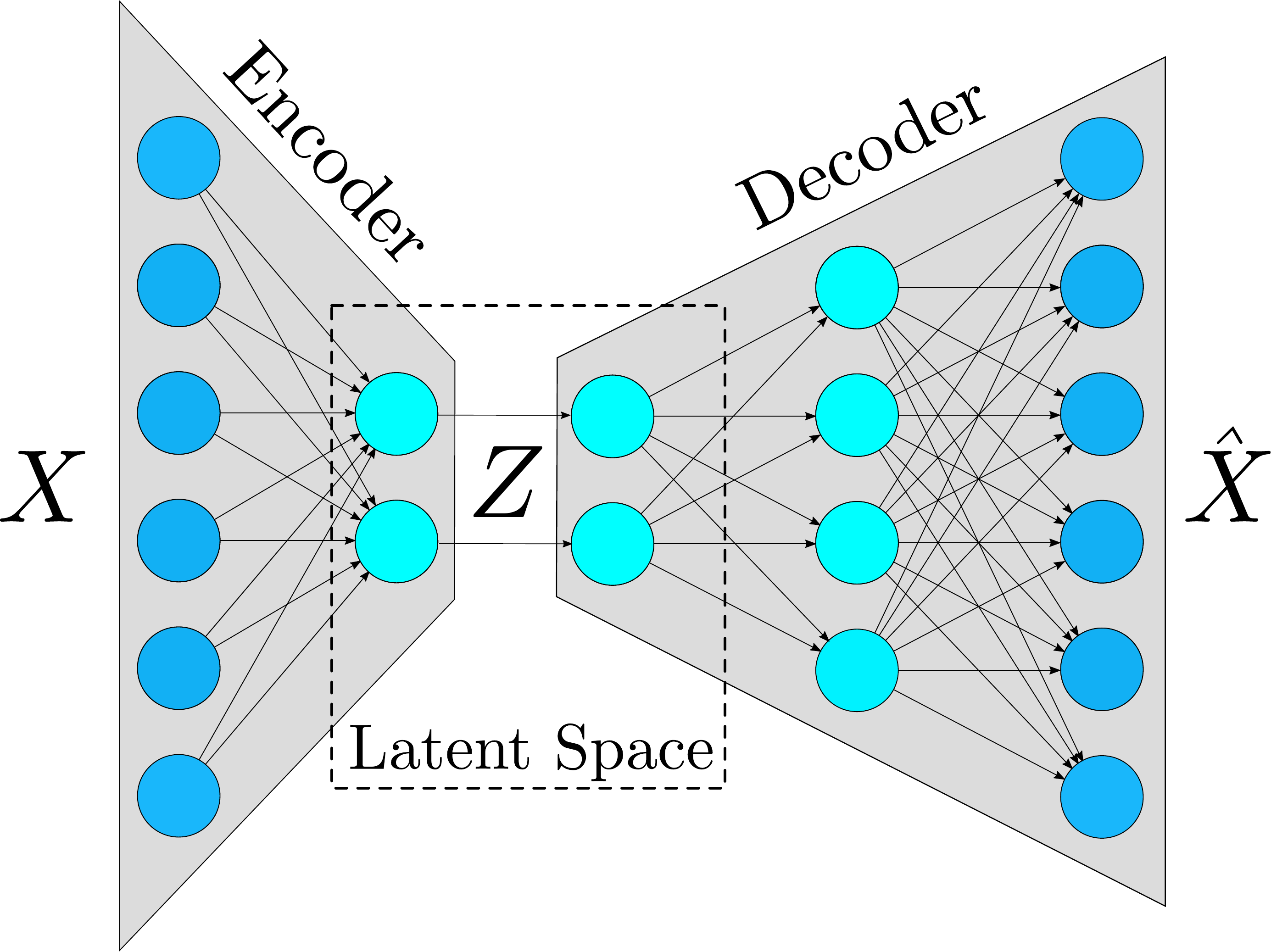}
  }
  \caption{Architecture of an autoencoder. The encoder transform the input data $X$ into a low-dimensional latent representation $Z$, while the decoder transforms the data back into the input space $\hat{X}$. For an optimal autoencoder it holds $X=\hat{X}$.}
  \label{fig:autoencoder}
\end{figure}

Autoencoders are a variant of ANN which is used for dimensionality reduction in a unsupervised learning setting~\cite{hinton1994autoencoders}.
The autoencoder consists of two parts: first an encoder, which transforms the high-dimensional input $X$ into a low-dimensional (\textit{latent}) representation $Z$.
This low-dimensional representation is then transformed back into the initial high-dimensional input space by the decoder, yielding $\hat{X}$.
The general layout is depicted in Fig.~\ref{fig:autoencoder}.
The characteristic hourglass shape forces the autoencoder to find an effective low-dimensional representation of the input.
The accuracy of the autoencoder can then be determined by computing the error of the network's input $X$ and output $\hat{X}$ vector.
Autoencoders are conceptually linked to the concepts of proper orthogonal decomposition (POD), principal component analysis (PCA) and singular value decomposition (SVD), since all of these concepts describe efficient low-dimensional representations of data. An extension of this method to variational autoencoders (VA) not only learns latent states $Z$, but a distribution given the input data. Thus, in inference mode, they are powerful tools for generative methods~\cite{kingma2013auto}, with application to the generation of synthetic turbulence in~\cite{fukami2018super,mohan2019compressed}.
\subsection{Reinforcement learning}
\label{sec:RL}
In contrast to supervised and unsupervised learning, the paradigm of reinforcement learning is not restricted to learn solely from given data points, but to learn from interaction with an environment and therefore to learn from experience.
The \textit{agent} is some machine learning algorithm that performs actions to interact with an environment.
These actions change the current state of the environment $s_{n}$ in a stochastic manner, i.e. the state change is generally not deterministic. 
Alongside the new state $s_{n+1}$, the agent receives a reward $r_{n+1}$ based on a reward function, which has to be adapted to the specific learning task.
The agent strives to adapt its actions $a$ in order to maximize the potential future award.
Training the agent then means to find a decision policy $\pi(a\:|s)$, which determines the actions the agent should take to maximize its future reward, given the current state.
The agent interacts with the environment across multiple~\textit{episodes} to gradually improve its policy $\pi$.
The most straight-forward approach is to explore all possible combinations of states $s$ and actions $a$, and memorize the expected reward in a table.
For future episodes, the agent can then lookup the current state $s$ and perform the action, which promises the highest reward.
This method is called Q-learning\cite{Watkins1992}.\\
Obviously this approach has limitations for large problems, since the memory cost and the time the agent needs to explore all possible combinations of states and actions is prohibitive for large problems.
Instead, the approach of deep reinforcement learning uses deep ANN to learn the policy for the RL algorithm, which was also used for AlphaGo, the first AI to beat a professional Go player \cite{Silver2016}.

\section{Examples of ML-augmented Turbulence modeling}\label{sec:turb}
In the previous sections, we have focused on the challenges and chances of ML methods in turbulence modeling and given a brief introduction to selected ML algorithms. In the following, we will discuss some successful applications of ML methods to RANS and LES problems in more detail. This discussion is far from complete, but we will try to match the presented cases to some of the challenges outlined above and show how these issues are tackled by the respective authors. We will focus here almost exclusively on supervised learning, mainly due to the much larger amount of research available than for other learning methods.\\
Recalling Sec.~\ref{subsec:why}, the task of turbulence modeling is to find a model $\hat{M}$ for the effect of the subfilter terms on the resolved solution. For the purpose of this overview, one possible hierarchy for ML-augmented turbulence modeling  is thus \emph{on which level} the modeling occurs: The first and most common approach is to replace the unclosed stresses by an a priori determined model function with parameters, and then fit or find the parameters through optimization approaches. This approach has the clear advantages that a) it can be used for model discrimination and that b) existing turbulence models are naturally incorporated. Further, the effect of the model onto the governing equations is typically via a modified viscosity, which is often very easy to incorporate into existing schemes. The main point here is to stress that the structural form of the model is posited a priori. We summarize these approaches as parameter estimation problems. Opposed to this, the second level of turbulence modeling directly predicts the closure terms (either the stresses or the forces), i.e. no functional form is assumed. The rationale behind this is to derive more universal closure models. The third level discussed here arguably leaves the realm of modeling - instead, it directly approximates the solution of the full (turbulent) equations through ML methods. While these methods give up some of the rigorous results of classical approximation schemes for PDEs, their motivation comes mainly from the fact that they are generally mesh-free and can work with sparse data.

\subsection{Parameter estimation}
LES or RANS models based on physical or mathematical considerations are typically formulated as (collection of) functions or transport equations of coarse grid data and associated tunable parameters. These parameters might be obvious, as a scaling factor of a viscosity for example, or more hidden in the details of the model and its implementation, as for the example the choice of where and what to average over in the dynamic Smagorinsky model. Since -- when coming from a turbulence modeling background -- parameter estimation tends to be seen as the obvious and most direct approach, a lot of literature focusing on this method with various flavors has been published in the last years. We do not aim to give a concise overview here, but highlight some interesting cases.\\
From the RANS modeling community, a number of publications on fitting the Reynolds stress tensor (or the discrepancy to existing models) based e.g. on a decomposition into its Eigenvectors have been published~\cite{tracey2013application,schmelzer2020discovery,wang2017physics}. Remarkable achievements have been the incorporation of field inversion techniques to reduce model-data inconsistencies and the direct embedding of constraints like Galilean invariances into the network~\cite{kutz2017deep,ling2016reynolds,parish2016paradigm}. An example from LES of a direct prediction of model coefficients is presented in~\cite{pal2019deep},where  the author learned the eddy-viscosity from a turbulent database, and showed good a priori and a posteriori LES results. The main feature of this work was the saving in computational time by a factor of 2 to 8 compared to an application of the dynamic Smagorinsky model. One of the earliest applications in learning a viscosity coefficient is due to Sarghini et al., who used an MLP to construct a more computationally efficient deep ANN
representation of Bardina's scale similarity model~\cite{sarghini2003neural}. For compressible turbulence, the authors in~\cite{xie2019artificial} propose to fit the coefficients of Clark's closure model through an NN. One fact that is shared by the LES-focused approaches in literature so far is that the question of interaction of discretization and model is often discarded by focusing either on explicitly filtered LES, or on global Fourier-based methods with a clear cut-off filter and homogeneous discretization properties.  Such practical difficulties are not (yet) included in the modeling process. An interesting exception to this for 2D turbulence is proposed by~\cite{maulik2020spatiotemporally}. Here, the implicit modeling method in LES is followed in the way that different numerical discretization operators for the inviscid fluxes (those with and without numerical dissipation) are chosen based on the classification by a neural network. The training data is obtained from a coarse-grained DNS field and is distributed into bins corresponding to regions of no, negative or positive eddy-viscosity. While combining numerical operators in such a way locally brings with it its own challenges, the results shown by the authors are promising and offer an interesting way of informing implicitly filtered, implicitly modeled LES. Care must be taken here in the future however to ensure the consistency of the filter used to classify the DNS data and the dissipative characteristics of the numerical operators.
\subsection{Closure term estimation}
\begin{figure}[htbp!]
	\centerline{
		\includegraphics[width=0.98\textwidth]{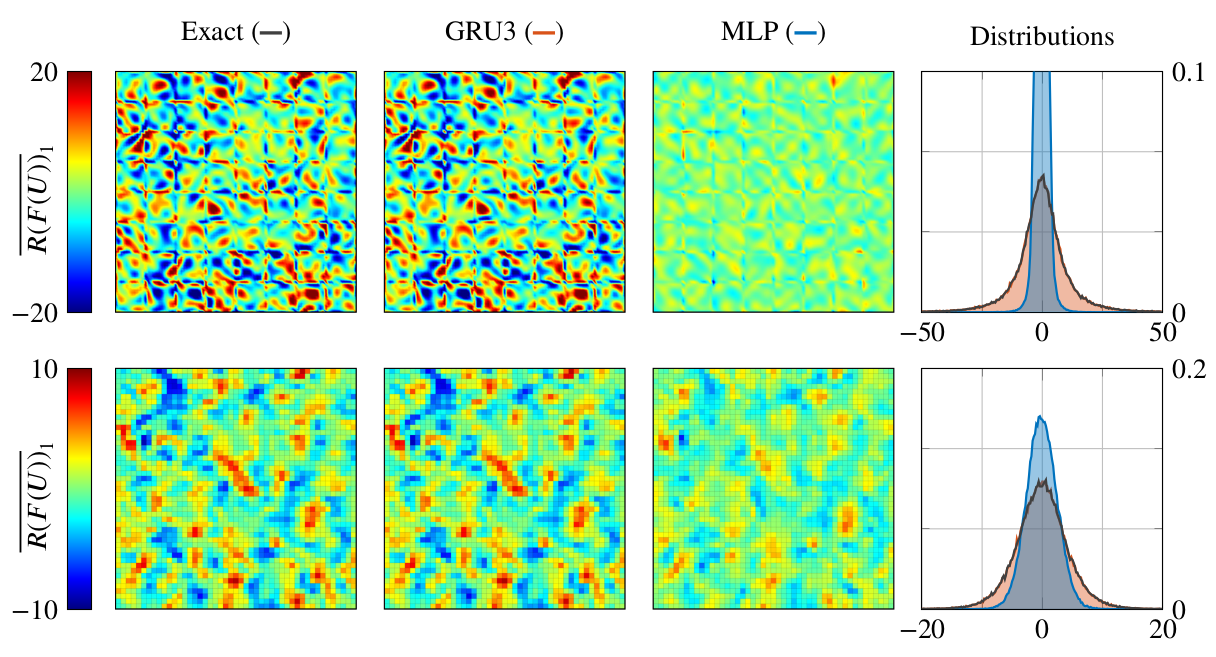}
	}
	\caption{Two-dimensional slices of the exact and predicted closure term of the x-momentum equation for a DHIT flow for two filters. Top row: Elementwise-$L_2$ projection onto the space of polynomials of degree $N=5$, bottom row: Top hat filter. The second column is obtained via a temporal NN based on a pointwise GRU network, the third column via a spatial MLP. Reproduced from~\cite{kurz2020machine}.}
	\label{fig:grupred}
\end{figure}

\begin{figure}[t]
	\centerline{
		\includegraphics[width=0.99\textwidth]{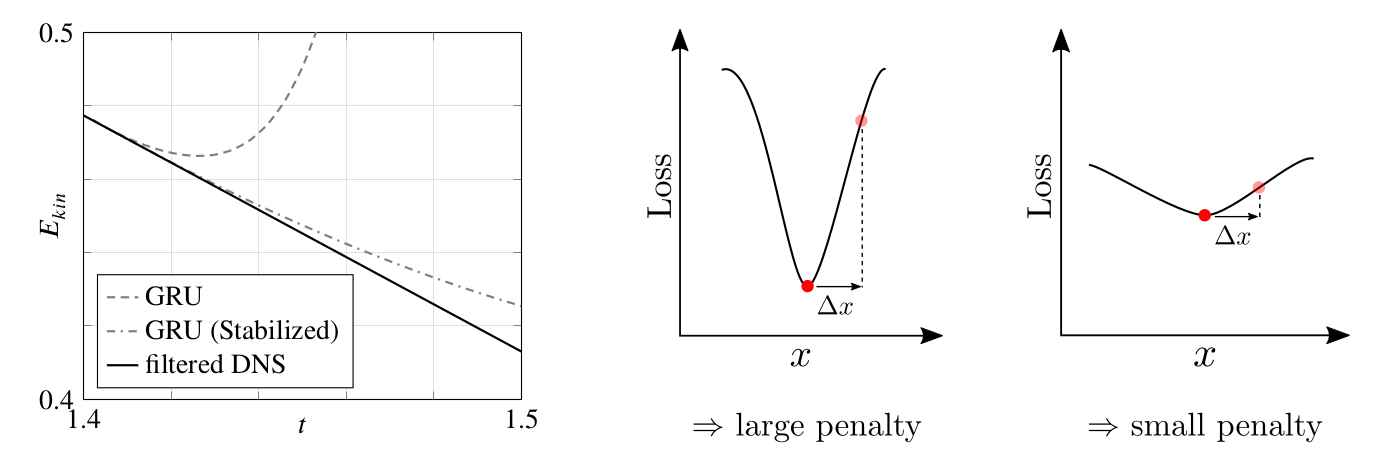}
	}
	\caption{\emph{Left:}~Evolution of the kinetic energy for the DHIT case. The filtered DNS data is given as reference. \emph{Right:}~Stability training to reduce the sensitivity to uncertain input data, basic idea from \cite{Zheng2016}. The objective of stability training is to prefer more shallow minima in the loss function to the more sharp ones during training, since the former generally result in better generalization properties. To determine whether the current minima of the loss function is shallow or sharp, random perturbations $\Delta x$ are applied to the input features. An additional penalty term is then added to the loss function, which is proportional to the induced changes in the loss function. To minimize the loss with the incorporated penalty term, the optimization procedure will tend to prefer the more shallow minima over the sharper ones and thus improve the generalization properties of the network.}
	\label{fig:ekin_stability}
\end{figure}

The direct prediction of the closure terms instead of their modeling offers an alternative to the parameter estimation task in the previous section. Here, the unknown terms in Eq.~\ref{eq:model} are directly approximated by the ML algorithm - either as fluxes or as the forces themselves~\cite{von2015towards,vollant2014optimal,vollant,beck2019deep,xie2020modeling,xie2019modeling,gamahara2017searching,ZHOU2019104319}. Important to stress however is the ambiguity of coarse to fine field, and thus of the coarse field to the closure terms. This one-to-many mapping is naturally introduced by the information loss in the LES subspace. For a given filter and thus coarse grained solution, an infinite number of associated fine scale fields and closure terms exist - thus, even exact closure terms should only be considered as means of the ensemble~\cite{moser2020statistical}.\\ 
The rationale for pursuing the direct estimation of the closure terms is three-fold: a) a direct closure avoids all modeling assumptions and is complete in a sense that it includes all the necessary and available information, b) if the perfect closure is available, this opens the door for fitting and developing more accurate models, c) the research question whether the closure terms can be recovered form coarse scale data only is of great interest in itself. In terms of the discussion in Sec.~\ref{subsec:why}, approaches from this group have some advantages: While they certainly need large amount of data to train on, the target quantity of the training is simply the filtered DNS data - no additional, derived quantities need to be available or computed. This reduces a strong source of data inconsistency and makes virtually any DNS data suitable for training. Also, finding the full closure terms instead of their models directly avoids any balancing issues between model terms, i.e. the optimization works on the level of the equations themselves. Physical constraints need still to be enforced, however, these constraints can directly be gleaned from the closure terms themselves and no interplay of different components of the model needs to be considered. A counterexample of this would be the scale-similarity model by Bardina with attached eddy-viscosity model, where the contributions of both model terms have to be balanced in this regard. There are of course also drawbacks to consider: interpretability of the resulting mapping is further reduced, as no functionals are posited a priori. Another, more severe shortcoming observed e.g. by ~\cite{vollant} is the lack of long-term stability in this form of closure. This stems from the unavoidable data-model inconsistency during inference as well as the error accumulation and self-driving error growth at high wavenumbers. This can either be tackled by removal of this energy through a dissipative mechanism~\cite{xie2020modeling} (essentially an additional model term) or the projection of the closure term onto a stable basis with the desired properties~\cite{beck2019deep}.\\
In general, the results for the closure term approximation based on ML methods are highly encouraging. In~\cite{xie2020modeling}, the authors used a spatial MLP with approx. 500,000 parameters to predict the three components of the subgrid forces of the incompressible momentum equations independently. As inputs, they chose velocity derivatives in the vicinity of the grid point in question. A priori correlations of over 90\% with the true subgrid force are reported by the authors. In an a posteriori application of the subgrid force together with a dissipative regularization term their approach outperforms classical closure models. In~\cite{von2015towards}, the authors applied a  method  for  a stochastic  model-discrimination based on so-called vector-valued auto-regressive models with external influences to the reconstruction of subfilter fluxes in Finite-Volume LES. Here, the model was not only able to capture the time-space structure of the subgrid fluxes reliably, but also to identify flow regimes, for example the near-wall region in a turbulent boundary layer that require their own local model (an example of an unsupervised clustering method). An approach to estimate the closure forces based on CNNs, which naturally incorporate spatial relationships in the mapping, was investigated in~\cite{beck2019deep}, with good success for the reconstruction. Additionally, the influence of an a priori feature selection revealed that the coarse-grained inviscid fluxes contribute noticeably to the training success. While the previous examples almost exclusively build their prediction from spatial data, in~\cite{kurz2020machine}, the temporal dimension was investigated as an input into NN-based approaches, specifically using the GRU architecture for sequential data (see Sec.~\ref{subsec:ann}). As an additional focus, the authors investigated the capabilities of the NNs to predict closure terms specific to a given filter kernel, an important capability of implicitly filtered LES. Fig~\ref{fig:grupred} presents the results of the predicted and true closure terms for a database of DHIT flows. As discussed in Sec.~\ref{subsec:turb}, the choice of the filter also defines the closure terms. For all filter types investigated, the GRU networks were able to achieve 99.9\% cross correlation in a priori tests based on a short series of temporal pointwise data of the coarse scale variables, making the prediction and target visually indistinguishable in the first two columns of Fig.~\ref{fig:grupred}. These results suggest that the subgrid force terms can be predicted with near arbitrary precision, which offers the chance to significantly improve practical closure models. As stated above, a direct closure with this predicted terms suffers from instability due to exponential error growth, and is thus not the method of choice without any regularization. This can come in the form of an improved model stability, or regularization during inference, for example through an additional dissipation mechanism. An example of the former approach is shown in the left diagram in Fig.~\ref{fig:ekin_stability}, where the evolution of the kinetic energy for the DHIT case is shown for GRU predictions. The filtered DNS results serve as a reference. While the GRU predictions lead to a very accurate closure and thus LES solution at first, the LES solution diverges strongly soon after. This is a result of a data-model inconsistency and the non-linear error accumulation of the truncated equations. One possible remedy is sketched in the right right of Fig.~\ref{fig:ekin_stability}, where so-called stability training was used successfully to flatten the cost function to reduce its sensitivity to uncertainties in the inputs during inference~\cite{Zheng2016}. This results in much greater stability of the closure model and longer useful predictions.

Summarizing this section, direct estimation of the subgrid forces has a range of advantages over parameter estimation approaches. However, without additional modeling or stabilization, it can lead to a diverging system in the long run. Nonetheless, availability of an accurate estimate of the exact closure at any time can likely help to improve classical modeling approaches. Currently, both spatial and temporal neural networks have been shown to be successful at this task, however, other approaches like kernel methods are likely also suitable. The generalizability of the results and the transfer of the predictions to a practical model are currently open questions. Partially related to the idea of closure term prediction is that of flow field deconvolution or approximate filter inversion, where the deconvolved field is then used to obtain (through scale similarity arguments) an approximation of the exact closure term. Inverting the filter by neural networks has for example been proposed in~\cite{maulik2017neural}, the reconstruction of the fine scale field through superresolution approaches (essentially the inversion of the cut-off filter) has been proposed in~\cite{fukami2018super}.
\subsection{Full PDE modeling}
\begin{figure}[t]
	\centerline{\includegraphics[width=0.75\textwidth]{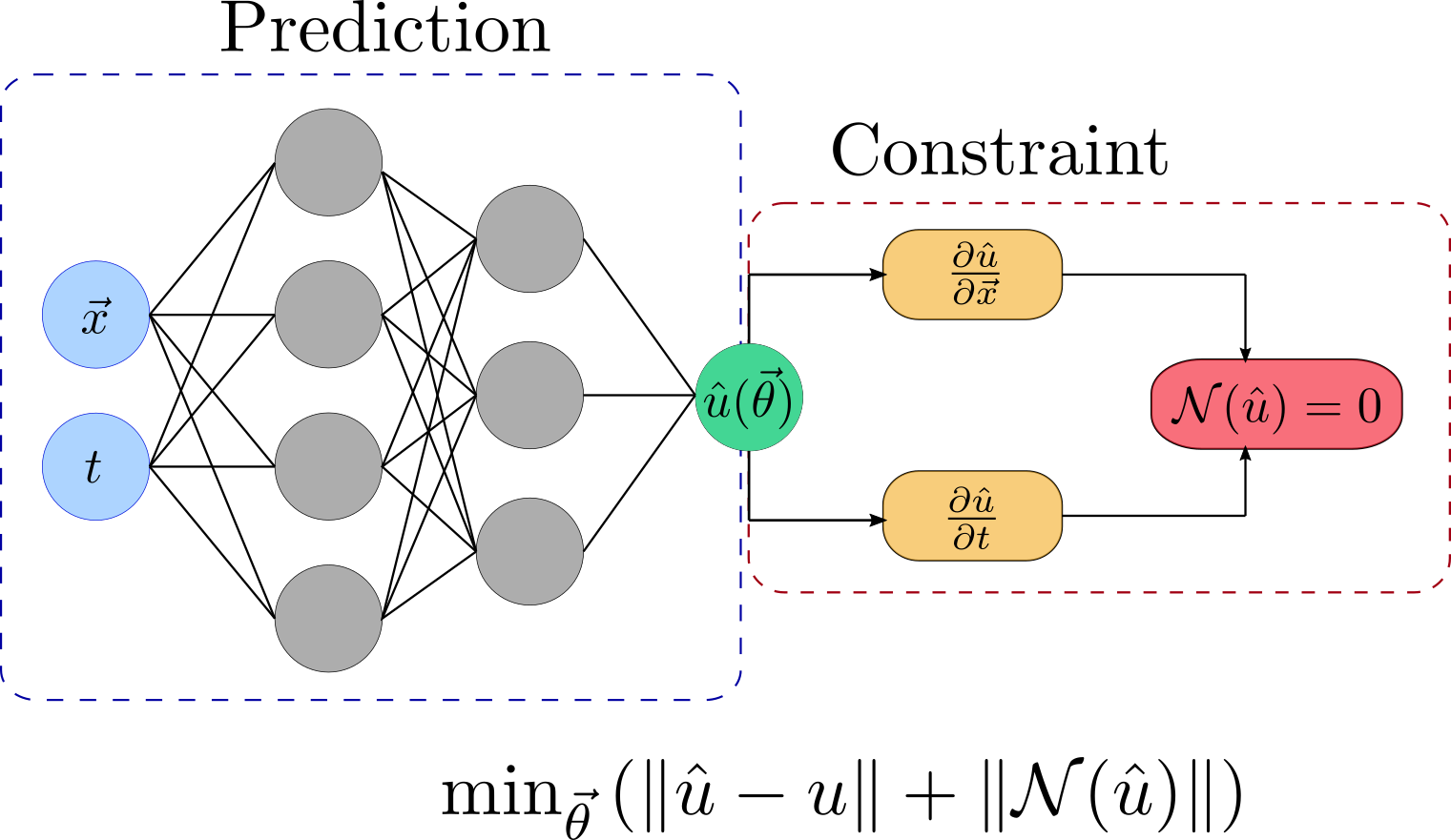}}
	\caption{A sample schematic of a PINN architecture for the linear transport equation $\frac{\partial u}{\partial t}+\frac{\partial u}{\partial \vec{x}}=0$.\label{fig:pinns}}
\end{figure}

In the rough hierarchy outlined above, the third level of ML-augmentation is the actual replacement of traditional PDE solution methods (i.e. the discretization of the underlying operators) by a learned approximation that is encapsulated in an ML algorithm. The so-called PINNs (Physics-informed neural networks) are a prominent approach that follows this idea. They can be understood as a combination of supervised (the prediction network) and unsupervised (the residual or constraint) learning methods. A schematic showing the general PINN idea is given in Fig.~\ref{fig:pinns} for the linear transport equation  $\mathcal{N}(u)=\frac{\partial u}{\partial t}+\frac{\partial u}{\partial \vec{x}}=0$. The task of the PINN is thus to find an approximation of $u$ under the constraints posed by the PDE. This is achieved in two consecutive steps: First, in a supervised manner, a predictor (some NN with parameters $\vec{\theta}$) for the solution field $\hat{u}(\vec{x},t,\vec{\theta})\approx u(\vec{x},t)$ is sought. The goal of this step is to obtain a parameterized function that represents $\hat{u}$ and that is expressed as a differentiable network graph. This graph, i.e. the prediction $\hat{u}$, can now be differentiated with automatic differentiation\footnote{Of course it could also be done analytically, but the graph itself is encapsulated in the network architecture and the framework, so automatic differentiation that is already available through back-propagation is a natural choice.} for any input feature, e.g. $\nicefrac{\partial \hat{u}}{\partial t}$. From these derivatives and other atomic operations, the original PDE can be reconstructed - it is of course a function of $\vec{\theta}$ and is thus optimizable by the ML method. Clearly, as the loss in the supervised part tends to zero, the residual will also tend to zero as a measure of how well the predicted solution fulfills the underlying PDE: $\hat{u}\rightarrow u \Rightarrow\mathcal{N}\rightarrow 0$. In this sense, the predicted solution can be conditioned on the PDE. The residual or constraint evaluation does not require any pre-labeled training data; the "correct" answer is given by the PDE itself. Thus, this part of the process can be seen as a unsupervised component. The optimization goal is then to find the best fit of both the solution field and the fulfillment of the PDE. Typically, in the PINN training process, the data on $u$ is given as boundary and initial conditions, while the sampling of the residual occurs at random points in feature space - it can of course also incorporate known solutions at given points. Additional constraints like periodicity can be included in a similar manner. \\
While a lot of work on PINNs for a range of problems exists~\cite{raissi2019physics,raissi2018hidden}, the application of PINNs to turbulence is still in its initial stages. In~\cite{jin2020nsfnets}, the authors present a PINN for the prediction of an incompressible laminar and turbulent flow in a DNS setting. Although the topic of this discussion is strictly speaking turbulence modeling, i.e. a coarse scale solution to the DNS, the PINN method can also be directly transferred to the LES or RANS equations. To our best knowledge, this has not been done yet, but would be a natural step. Nonetheless, PINNs have some compelling properties that could be exploited for RANS and LES and are thus discussed here alongside the other approaches.\\
 The training samples used in~\cite{jin2020nsfnets} for the prediction network come from analytical or DNS boundary data, while the residual loss is evaluated on points scattered in the feature domain. The PINN predictions for laminar flow achieve low approximation errors, while these errors become considerably larger in the turbulent case. The authors report in general a good agreement of short term predictions, but data indicates that error accumulates in the long run. An interesting outcome of their investigations is that although no training data on the pressure field is provided, the enforcement of the divergence condition through the constraint-part of the network leads to the expected pressure field, without having to split the equations. It is also remarkable that the networks for laminar flows are much smaller than those for the turbulent cases (all are MLPs), with the largest one having approx. $9\times 300^2 \approx 800,000$ learnable parameters. The efficiency of this approach compared to the classical PDE solution strategies is an open issue. Also, although the authors note that their approach does not require grids nor does it introduce the classical dispersion and diffusion errors, it remains clear that the approach is not errorfree and remains an approximation to the true PDE solution. This entails that properties built into classical approximation schemes like conservation are also only approximated by PINNs, and that a mathematical analysis of the approach in terms of stability and convergence behavior is an open field of research.\\ 
In light of the discussion in Sec.~\ref{subsec:why}, PINNs and associated methods offer two desirable features for turbulence modeling or turbulence simulation: As the equations themselves are approximated, also all inherent physical constraints are naturally included. In addition, PINNs exploit the equations themselves and thus need very little labeled training data - the vast majority of training data can be unsupervised and thus generated on the fly. This also removes the typical source of training data inconsistency. In addition, the data can be scattered randomly or focused in regions of high solution variance and the approach allows the direct inclusion of available DNS or experimental data. In how far a trained PINN is able to generalize is an issue that needs to be explored further, also, work on interpretability is generally as incomplete as it is for other NN-based approaches. Finally, the question of efficiency arises: can a PINN be faster than a traditional PDE solver? If so, at what accuracy? Is there a hierarchy of approximations in the sense that a PINN for an LES or RANS solution can indeed be much smaller and thus faster than that for a DNS? Can they be stable if the underlying PDE is in itself an approximate model?\\
 These open questions indicate that there is a lot of research necessary to ascertain the place of PINNs in turbulence modeling and turbulence simulation. In particular, an issue of debate will be the new approximation paradigm introduced by PINNs in that it requires (at least for now) relaxing many of the known and relied-upon properties of classical PDE solvers. However, the advantages of PINNs have also been demonstrated by their original authors; thus, where and how PINNs can fit will have to be established through future research.

\section{Conclusion and outlook}\label{sec:outro}
Progress in turbulence modeling has been rather incremental in the last decades, with a lot of efforts focused on improving existing models in terms of their range of application, generalizability, stability and performance when combined with different numerical schemes. The development of new models has so far predominantly been driven by mathematical and physical considerations. Data-driven approaches have recently gained a strong interest and have been proposed as a "third way" in closure modeling, with the potential of considerably accelerating and possibly unifying model development. \\
From the authors' perspective, ML-augmented modeling will make here neither physical reasoning nor mathematical rigor obsolete. In fact, the opposite is even likely. As discussed in this paper, ML does not necessarily make model development easier or more fool-proof, instead, its successful application requires a careful appreciation of the underlying equations, their properties, their discretizations and all the intricate and sometimes obscure details of the LES and RANS methods. In fact, the focus on ML methods might bring an old but often overlooked inconsistency in LES into focus: The separation of a model from the mathematical and physical environment and conditions it was derived for. Based on the theoretical work on explicitly filtered LES and "perfect" numerical schemes, it is easy to subscribe to the notion that there exists e.g. one Smagorinsky model, and that its associated constant should thus be - based on physical consideration - universal. However, in practical, grid-filtered LES computed with imperfect discretization operators, many variants of the Smagorinsky formulation exist (e.g. depending on how the associated velocity gradients are computed), and the considered model interacts with this "dirty" data and environment in complex forms. This issue contributes to large variations reported in LES benchmark results.\\
ML methods can bring these model-data-inconsistencies into sharper focus, and a lot of research efforts are underway to account for them. Thus, a success in ML-augmented turbulence modeling will likely also mean a more thorough understanding and appreciation of the other aspects of LES, RANS and turbulence, and help to make the definition of what we can expect from a coarse-grain solution sharper. In this sense, ML can offer capabilities that are indeed complementary to more traditional approaches. From our perspective, the feature extraction or the identification of undiscovered correlations from data are among the most important ones. \\
The initial successes of ML in improving existing models or reconstructing the closure terms directly are remarkable. What must come next is a consolidation of the many research directions, and a generalization of the models to make them actually useful in real applications. This task is made rather difficult by the fact that although the initial driving force behind data-augmentation in turbulence modeling might have been the desire to derive more universal models and reduce empiricism, ML introduces its own new level of hyperparameters, methods and uncertainties. The resulting ML models can have millions of tuneable parameters, and may lose all interpretability. Even transferring a published model from one researcher to the next becomes not merely the task of providing a set of equations, but the exchange of graphs, training environments and possibly large amounts of data. Another unanswered issue is the cost-benefit trade-off, both during training as well as during inference.\\ Thus, in order to achieve a level of consolidation of the research efforts and the chance to select the most promising approaches to go forward, the next step should be the establishment of a means of comparing and benchmarking the results. Such a database should contain a well-described training dataset, most likely from DNS, open to everyone, and a private test set for the blind evaluation of the proposed models. Setting up such a validation case and administering it should be approached as a community effort.

\section*{Acknowledgments}
This research was funded by Deutsche Forschungsgemeinschaft (DFG, German ResearchFoundation) under Germany's Excellence Strategy - EXC 2075 - 390740016.
The authors gratefully acknowledge the support and the computing time provided by the HLRS through the project ``hpcdg''.

\bibliography{main.bib}{}
\bibliographystyle{siamplain}

\end{document}